\documentclass{article}
\usepackage{amssymb}

\usepackage{graphicx}
\usepackage{amsmath}

\input{tcilatex}
\usepackage{graphicx}

\begin{document}

\title{Statistical Mechanics of 2+1 Gravity From Riemann Zeta Function and
Alexander Polynomial: Exact Results}
\author{A.L.Kholodenko \\
%EndAName
375 H.L.Hunter Laboratories, \\
Clemson University, Clemson, \\
SC 29634-0973, USA}
\maketitle

\begin{abstract}
In the recent publication (Journal of Geometry and Physics, 33 (2000) 23-102
) we have demonstrated that dynamics of 2+1 gravity can be described in
terms of train tracks. Train tracks were introduced by Thurston in
connection with description of dynamics of surface automorphisms. In this
work we provide an example of utilization of general formalism developed
earlier. The complete exact solution of the model problem describing \
equilibrium dynamics of train tracks on the punctured torus is obtained.
Being guided by similarities between the dynamics of 2d liquid crystals and
2+1 gravity the partition function for gravity is mapped into that for the
Farey spin chain. The Farey spin chain partition function, fortunately, is
known exactly and has been thoroughly investigated recently. Accordingly,
the transition between the pseudo-Anosov and the periodic dynamic regime (in
Thurston's terminology) in the case of gravity is being reinterpreted in
terms of phase transitions in the Farey spin chain whose partition function
is just the ratio of two Riemann zeta functions. The mapping into the spin
chain is facilitated by recognition of a special role of the Alexander
polynomial for knots/links in study of dynamics of self homeomorphisms of
surfaces. At the end of paper, using some facts from the theory of
arithmetic hyperbolic 3-manifolds ( initiated by Bianchi in 1892), we
develop systematic extension of the obtained results to noncompact Riemann
surfaces of higher genus. Some of the obtained results are also useful for
3+1 gravity. In particular, using the theorem of Margulis, we provide new
reasons for the black hole existence in the Universe: black holes make our
Universe arithmetic. That is the discrete Lie groups of motion are
arithmetic.
\end{abstract}

\section{ Introduction and summary}

\noindent

The Riemann zeta function $\zeta(\beta)$ has been an object of intensive
study in both mathematics [1-3] and physics [4,5] for quite some time.The
reason for physicists interest in this function can be easily understood if
one writes it in the form of a partition function Z($\beta)$ given by 
\begin{equation}
\zeta(\beta)\equiv Z(\beta)=\sum\limits_{n=1}^{\infty}\exp\{-\beta\ln n\}.  
\tag{1.1}
\end{equation}
If $\beta$ is interpreted as the inverse temperature then, naturally,
questions arise:

a) what is the explicit form of the quantum mechanical Hamiltonian whose
eigenvalues $E_{n}$ are given by $E_{n}=\ln n$ ?

b)can such system undergo phase transition(s) if one varies the temperature?
\ \ \ \ \ \ \ \ \ \ \ \ \ \ \ \ \ The goal of providing answers to both
questions is at the forefront of current research activities both in physics
[4,5] and mathematics [6]. Answers to these questions are being sought in
connection with theories of random matrices and quantum chaos [4,5],
non-commutative geometry [6] and Yang-Lee zeros [7]. According to the theory
of Yang and and Lee the problem of existence of phase transitions can be
reduced to the problem of existence of zeros of the partition function in
the complex z-plane (where z may be related to either fugacity or the
magnetic field, etc.). In the case of $Z$($\beta)$, Eq.(1.1), one is also
looking at analytic behavior of the partition function in the complex $\beta-
$plane. Riemann had conjectured that, 
\begin{equation}
Z(\frac{1}{2}+it_{m})=0,   \tag{1.2}
\end{equation}
provided that $Re$ $t_{m}\neq0$ for all integer $m$'s. This conjecture is
known in the literature as Riemann hypothesis. Stated differently, the
Riemann hypothesis is equivalent to the statement that \ all ''nontrivial''
zeros of the partition function Z($\beta)$ are located at the critical line $%
Re\beta=\frac{1}{2}.$ The ''trivial'' zeros are known [1-3] to be located at 
$\beta=-2,-4,-6,....,$that is, 
\begin{align}
Z(\beta & =-2m)=0,  \tag{1.3} \\
Z(\beta & =-2m+1)=-\frac{\text{B}_{2m}}{2m}  \notag
\end{align}
for $m$=1,2,... and B$_{2m}$ being the Bernoulli numbers.

Combinations of the Riemann zeta functions are also of physical interest. In
particular, in this paper we shall be concerned with the following
combination 
\begin{equation}
\hat{Z}(\beta)=\frac{\zeta(\beta-1)}{\zeta(\beta)}=\sum\limits_{n=1}^{\infty
}\phi(n)n^{-\beta}   \tag{1.4}
\end{equation}
where $\phi(n)$ is the Euler totient function, 
\begin{equation}
\phi(n)=n(1-\frac{1}{p_{1}})\cdot\cdot\cdot(1-\frac{1}{p_{r}}),   \tag{1.5}
\end{equation}
which is just the number of numbers less than n and prime to $n$, provided
that $n=p_{1}^{m_{1}}\cdot\cdot\cdot p_{r}^{m_{r}}$ , and $p_{1}$, etc. are
primes with respect to $n$. This partition function had appeared in
mathematical physics literature in connection with the partition function
for the number-theoretic spin chain [8-10] and in connection with
calculations of the scattering S-matrix for the ''leaky torus'' quantum
mechanical problem [11-13]. Remarkably enough, the results for the
number-theoretic spin chain can be obtained as well from earlier works on
mode locking and circle maps [14,15], e.g. see in particular Eq.(30) of
Ref.[15], as the authors of Ref.[10] acknowledge. This fact is not totally
coincidental as we shall explain below (in sections 2 and 3) and has been
already anticipated based on our earlier works [16,17] on dynamics of 2+1
gravity.

In this paper we would like to demonstrate that the partition function,
Eq.(1.4), can adequately describe statistical mechanics of Einsteinian 2+1
gravity if the underlying surface is the punctured torus. The restriction to
the punctured torus case is not too severe and is motivated mainly by
illustrative purposes: recall, that both the Seifert surfaces of the figure
eight and the trefoil knots are just punctured toruses [18]. This
observation allows us to make an easy connection between the dynamics of
surface self homeomorphisms and the associated with its time evolution
3-manifolds which fiber over the circle [17] . These manifolds are just
complements of the figure eight and trefoil knots in S$^{3}$
respectively.The Seifert surfaces of more complicated knots may naturally be
of higher genus but, since both the trefoil and the figure knots belong to
the category of fibered knots, only those knots and links which are fibered
and the associated with them Seifert surfaces are relevant to the dynamics
of 2+1 gravity [17]. The 3- manifolds associated with the figure eight and
the trefoil knots are fundamentally different: the first one is known to
belong to the simplest representative of the hyperbolic manifolds while the
second corresponds to the so called Seifert fibered manifolds [19].The
surface dynamics associated with the first is associated with pseudo-Anosov
type of surface self homeomorphisms while the second one is associated with
periodic self homeomorphisms. Both types of 3-manifolds are topologically
very interesting and potentially contain wealth of useful physical
information. In this work we only initiate their study with hope of
returning to this subject in future publications.

In order for the reader to keep focus primarily on physical aspects of the
problem, we feel, that some simple explanation of what follows is
appropriate at this point. To avoid repetitions, we expect that our readers
have some background knowledge of the results presented in our earlier
published papers [16,17]. In particular, to help our intuition, we would
like to exploit the fact that statics and dynamics of 2+1 gravity is
isomorphic with statics and dynamics of textures in two dimensional liquid
crystals. According to the existing literature on liquid crystals, e.g.
[20], the liquid crystalline state can be found in several phases which
physicists classify as liquid, solid, hexatic and gas. In mathematical
literature the textures, e.g. like those in liquid crystals, are known as
foliations [19,21]. Dynamics of textures is known accordingly as dynamics of
foliations. Some of these foliations may contain singularities.These
singularities are mistakenly being treated as Coulombic charges (while in
2+1 gravity is is well documented [17,22] that these singularities do not
interact) since the \textbf{nonorientable} line fields are being confused
with the \textbf{orientable} vector fields. The phase transitions in two
dimensional liquid crystals are described in terms of the phase transitions
in 2 dimensional Coulomb gas. These are known as the Kosterlitz-Thouless
type transitions [23]. Although mathematically such explanation of phase
transition is not satisfactory, nevertheless, one has to respect the
experimental data associated with the liquid crystalline phases.The extent
to which the Kosterlitz-Thouless interprepretation of phase transition is
appropriate is discussed in the Appendix A.1. where it is being argued that,
by analogy with transitions in the liquid helium (where the
Kosterlitz-Thouless interpretation is normally used) the dynamical phase
transition in 2+1 gravity resemble to some extent the Bose gas condensation
type of transition.This analogy is incomplete, however, and is only being
used for the sake of comparison with the existing literature. As Remark 4.1.
indicates, the partition function of 2+1 gravity \textbf{without any
appriximations} can be recast into the Lee-Yang form [7]. The calculations
associated with such form require knowledge of distribution of zeros of the
Riemann Zeta function and, hence, effectively, the proof of the Riemann
hypothesis. Since this proof is not yet available (see, however, Ref.[5]) we
employ alternative methods associated with recently developed
thermodynamic/statistical mechanic formalism for description of phase
transitions in the number-theoretic Farey spin chains [8-10].\ To prove that
dynamical transitions in gravity can be described in terms of transitions in
the Farey spin chains several steps are required. In our earlier works
[16,17] we have demonstrated that dynamic of 2+1 gravity is best described
in terms of dynamic of train tracks. In section 2 we demonstrate how dynamic
of train tracks can be mapped into dynamic of geodesic laminations. In turn,
the dynamic of geodesic laminations is reformulated in terms of the sequence
of Dehn twists. This sequence is actually responsible for the \textbf{%
dynamic in the Teichm\"{u}ller space} of the punctured torus. Such dynamic
is subject to the number-theoretic constraints associated with the Markov
triples.The Markov triples had been known in physics literature for a while
in connection with the trace maps [24] used for description of
quasicrystals, 1d tight band Scr\"{o}dinger equations, etc. In this work the
Markov triples play a somewhat different role: they make the set of closed
nonperipheral geodesics on the punctured torus discrete. Mathematically, the
sequence of Dehn twists is written in terms of the product of the ''right''
and the ''left'' 2$\times2$ Dehn matrices. The modulus of eigenvalues(the
stretch factors) associated with such matrix product can be either greater
than one \ or equal to one. In the first case one is dealing with the
pseudo-Anosov and in the second, with the periodic (Seifert-fibered)
dynamical regime. The results of section 2 (for the figure 8 and the trefoil
knots) acquire new meaning in section 3 where they are reobtained with help
of the associated Alexander polynomials.The stretch factors of section 2 are
reobtained as zeros of the related Alexander polynomials. In the same
section 3 we discuss the fiber bundle construction of 3-manifolds
complementary to the figure eight and the trefoil knots in $S^{3}.$ The
sequence of Dehn twists, discussed in section 2, in this section is being
associated with the operation of Dehn surgery (Dehn filing) performed on the
3-manifold related to the figure eight knot.The stretch factors produced as
result of such surgery are reobtained with help of the Mahler measures which
allow us to reinterpret these factors in terms of the topological entropies.
Introduction of the Mahler measures, in addition, allows us to make direct
connection between the dynamical phase transitions and the thermodynamic
transitions in the sense of Yang and Lee [7]: zeros of the Alexander
polynomial play similar role in dynamics as Yang-Lee zeros in
thermodynamics. In section 4 we provide direct connection between the
results of section 3 just described and the statistical mechanics formalism
developed for the number-theoretic Farey spin chains [8-10]. With such
connection established, dynamical phase transitions in 2 +1 gravity can be
treated with formalism which is more familiar to physicists. Unfortunately,
this more familiar formalism is applicable (at least at the present level of
our understanding) only to the case of punctured torus. Fortunately, the
final results obtained with help of such type of formalism can be reobtained
in several different and independent ways.We discuss these alternative ways
at the end of section 4 and in section 5 which is entirely devoted to
development and refinements of these alternatives. The major reason of doing
this lies in the opportunity of extension of the punctured torus results to
the noncompact Riemann surfaces of \textbf{any} genus. This is accomplished
using some results from the scattering theory for Poincar$e^{\prime}$
(Eisenstein) series acting on 3-manifolds. These series had been recently
discussed in our earlier work, Ref.[25], to which we refer for more details.
The key result of section 4, the partition function for the case of
punctured torus, happens to coincide with the scattering S-matrix (up to
unimportant constant) obtained for 3-manifolds with one \textbf{Z}$\oplus%
\mathbf{Z}$ cusp [26]. 3-manifolds containing multiple \textbf{Z}$\oplus%
\mathbf{Z}$ cusps are associated with fibered links (as explained in section
3 and the Appendix A.3). The S-matrix for this case has been also obtained
recently [27]. The determinant of this matrix produces the desired exact
partition function for 2+1 gravity. In addition, the formalism allows us to
obtain the volumes of the associated 3-manifolds exactly. Extension of the
punctured torus results to surfaces of higher genus requires some careful
analysis of the nontrivial mathematical problem of circle packing. This fact
has some profound impact at development of the whole formalism. It happens,
that such type of problems had been comprehensively studied by Bianchi
already in 1892 [28] who championed study of the arithmetic hyperbolic
manifolds. The notion of arithmeticity is rather involved. Since in physics
literature (to our knowlege) it did not find its place yet, in Appendices
A.2 and A.3 we supply the essentials needed for the uninterrupted reading of
the main text. Surely, selection of the material in these appendices is
subjective. But, it is hoped, that interested reader will be able to restore
the missing details if \ it is required. The arithmeticity of 3-manifolds
associated with 2+1 gravity \ stems from some very deep results of Riley
[29] and Margulis [30] which we discuss to some extent in Appendix A.3. The
arithmeticity leads to some restrictions on groups of motions in symmetric
spaces (e.g. hyperbolic space is symmetric space). Thanks to the Margulis
Theorem A.3.11. and some results of Helgason [31] and Besse [32], the notion
of arithmeticity is extendable to 3+1 gravity as well. In the case of 2+1
gravity we demonstrate, that the very existence of black holes makes such
2+1 Universe arithmetic. Since most of the Einstein spaces happen to be
symmetric, we expect that they are arithmetic in addition in view of the
Margulis theorem [30].This possibility is realized in nature only if the
black holes exist in our Universe. The black holes make our Universe
arithmetic.

\section{From train tracks to geodesic laminations}

\subsection{\protect\bigskip Dynamics of train tracks on punctured torus}

Dynamics of pseudo-Anosov homeomorphisms on the four punctured sphere \ was
studied in some detail in Ref.[33].Closely related but more physically
interesting case is associated with study of self homeomorphisms of the
punctured torus.The Poincare-Hopf index theorem requires existence of two
Y-type singularities, each having index -$\frac{1}{2}$, as it is explained
in Ref.[16]. These singularities can move on the surface of the punctured
torus thus giving rise to the train tracks dynamics as depicted in Fig.1%
\textbf{.\FRAME{ftbpFU}{4.6267in}{5.2312in}{0pt}{\Qcb{Fragment of the train
track dynamics on once punctured torus}}{}{fig.1.gif}{\special{ language
"Scientific Word"; type "GRAPHIC"; maintain-aspect-ratio TRUE; display
"USEDEF"; valid_file "F"; width 4.6267in; height 5.2312in; depth 0pt;
original-width 9.1982in; original-height 10.4063in; cropleft "0"; croptop
"1"; cropright "1"; cropbottom "0"; filename 'C:/New
Folder/fig.1.gif';file-properties "XNPEU";}}}

From this picture it follows, that the nontrivial dynamics is effectively
caused by sequence of meridional $\tau_{m}$ and longitudinal $\tau_{l}$ Dehn
twists. Using the rules set up for dynamics of train tracks [16], one can
easily calculate the transition matrix by noticing that topologically the
state h) is the same as a) while the weights on the branches are different.
This allows us to write the following system of equations 
\begin{equation*}
a^{\prime}=b+2a, 
\end{equation*}
\begin{equation}
b^{\prime}=c,   \tag{2.1}
\end{equation}
\begin{equation*}
c^{\prime}=a+2c. 
\end{equation*}
These results can be neatly presented in the matrix form : 
\begin{equation}
\left( 
\begin{array}{c}
a^{\prime} \\ 
b^{\prime} \\ 
c^{\prime}
\end{array}
\right) =\left( 
\begin{array}{ccc}
2 & 1 & 0 \\ 
0 & 0 & 1 \\ 
1 & 0 & 2
\end{array}
\right) \left( 
\begin{array}{c}
a \\ 
b \\ 
c
\end{array}
\right) .   \tag{2.2}
\end{equation}
The incidence matrix just obtained can be found in Penner's article,
Ref.[34], where it was presented without derivation.The largest eigenvalue $%
\lambda$ is found to be 
\begin{equation}
\lambda=\frac{1}{2}(3+\sqrt{5}).   \tag{2.3}
\end{equation}
Since $\lambda>1,$this indicates that the dynamics depicted in Fig.1 is of
pseudo-Anosov type. We shall reobtain this result for $\lambda$ below, e.g.
see Eq.(2.16), and in section 3 using totally different methods. In fact,
the presentation above is only given for the sake of comparison with these
new methods to be discussed in section 3. These new methods provide the most
natural connections between the dynamics of 2+1 gravity and the theory of
knots and 3-manifolds.

The reminder of this section is devoted to exposition of some mathematical
results which will be used in the rest of this paper.

\subsection{The Markov triples}

To begin, let us recall [16,17,19], that a \textit{geodesic lamination} on a
hyperbolic surface S is a closed subset of S made of union of disjoint
simple geodesics. When lifted to the universal cover, i.e. to the Poincar$%
e^{\prime}$ disc $\mathcal{D}$ whose boundary is a circle $S_{\infty}^{1}$
at infinity, the endpoints of the geodesic lamination determine a closed
subset (actually, a Mobius strip) 
\begin{equation}
\mathcal{E}=(S_{\infty}^{1}\times S_{\infty}^{1}-\Delta)/Z_{2}   \tag{2.4}
\end{equation}
where $\Delta$ is diagonal $(x,x),x\in S_{\infty}^{1}$ and the factor $Z_{2}$
reflects the fact that the circle segments representing these geodesics are
unoriented, that is the picture remains unchanged if the ends of each
geodesic which lie on $S_{\infty}^{1}$ are interchanged. This fact has been
discussed and used already in our earlier work, Ref.[16]. Since, by
definition, the lamination is made of disjoint set of geodesics, when lifted
to $\mathcal{D}$, there are no circle segments (representing geodesics)
which intersect with each other. It is intuitively clear, that the dynamics
of train tracks should affect the dynamics of geodesics. We would like to
make this intuitive statement more precise. To this purpose, we notice that,
when lifted to the universal cover, this dynamics causes some
homeo(diffeo)morphisms of the circle $S_{\infty}^{1}$ thus making clear the
connections with circle maps and mode locking [14,15].

If $G=\pi_{1}(S)$ \ is the fundamental group of surface $S$, the endpoint
subset $\mathcal{E}$ remains invariant under the action of $\pi_{1}(S)$.
Suppose, that the surface $S$ has boundaries, e.g. a hole in the case of a
torus. Let $\mathcal{\ }P\sqsubseteq G$ be the set of \emph{peripheral
elements},e.g. those elements of $G$ which correspond to loops freely
homotopic to the boundary components. In the case of a punctured torus, $G$
is just a free group of two generators: $G=<a,b>,$ and the subset $P$ is
determined by the commutator $aba^{-1}b^{-1}\equiv\lbrack a,b]$ whose trace
equals to -2 [35]. Such restriction on the trace indicates that the group
element $[a,b]$ is parabolic.Parabolic elements are always associated with
cusps (punctures) on the Riemann surface (for more details, e.g.consult
Refs.[25,29]).This restriction affects the presentation of the group $G.$
Using generators $a$ and $b$ one can construct a ''word'' $W_{r}$%
\begin{equation}
W_{r}=a^{\alpha_{1}}b^{\beta_{1}}\cdot\cdot\cdot a^{\alpha_{r}}b^{\beta_{r}} 
\tag{2.5}
\end{equation}
where $\alpha_{1}$ and $\beta_{r}$ can be any integers while $\alpha_{i}$
and $\beta_{i\text{ }}$, $i\neq1,r,$ can be any integers, except zero. Not
all words thus constructed are different.Those words which are conjugate,
i.e. $W_{r}W_{l}=W_{l}W_{r}$ , are considered \ to be equivalent. If we
define the equivalence relation as $\sim$ , then the quotient $G/\sim$ can
be identified with the set $\Omega$ of homotopy classes of closed curves on
the torus T [35]. Let $\hat{\Omega}$ be a subset of $\Omega$ which
corresponds to the nontrivial nonperipheral simple closed curves on T . To
account for the presence of a puncture, we require that \textbf{both} $a$
and $b$ belong to $\hat{\Omega}$ while $[a,b]$ should correspond to the
peripheral simple closed curve(s).The question now arises: how to find an
explicit form of the generators $a$ and $b$? If we require these generators
to act by isometries in the Poincare disc $\mathcal{D}$ (or the upper half
plane ) we need to find a mapping of the group $G$ into the group PSL(2,R).
This can be achieved with help of the following identities (in the case of
SL(2,R)): 
\begin{equation}
2+tr[a,b]=\left( tra\right) ^{2}+\left( trb\right) ^{2}+\left( trab\right)
^{2}-tra\text{ }trb\text{ }trab   \tag{2.6a}
\end{equation}
and 
\begin{equation}
tra\text{ }trb=trab+trab^{-1}   \tag{2.6b}
\end{equation}
which had been discovered by Fricke [36].The above identities can be
analytically extended to SL(2,C) [35]. The projectivized versions of these
groups, that is PSL(2,R) and PSL(2,C), are groups of isometries of
hyperbolic H$^{2}$ and H$^{3}$ spaces respectively. \ Such an extension is
useful for dealing with problems discussed in our earlier work, Ref.[25],
and will be also discussed later in this work in section 5.

Since, as we know already, $tr[a,b]=-2$, the above identities can be
conveniently rewritten as 
\begin{equation}
x^{2}+y^{2}+z^{2}=xyz   \tag{2.7}
\end{equation}
and 
\begin{equation}
xy=z+w,   \tag{2.8}
\end{equation}
where, $x=tra,y=trb,z=trab$ and $w=trab^{-1}.$ As it is argued by Bowdich in
Ref.[37] (who, in turn, attributes it to Jorgensen [38]) the first identity,
Eq.(2.7), is sufficient for restoration of the explicit form of matrices $a$
and $b$. These are given by 
\begin{equation}
a=\frac{1}{z}\left( 
\begin{array}{cc}
xz-y & x \\ 
x & y
\end{array}
\right) \text{ and }b=\frac{1}{z}\left( 
\begin{array}{cc}
yz-x & -y \\ 
-y & x
\end{array}
\right) .   \tag{2.9a}
\end{equation}
The above matrices differ slightly from those given in Ref.[37]. This
difference is essential, however, and originates from the fact that we
require $deta=detb=1.$ Such requirement leads automatically to Eq.(2.7) as
required. The above choice of matrices is not unique since, for example,
Eq.(2.7) is also to be satisfied by the choice of 
\begin{equation}
a=\frac{1}{z}\left( 
\begin{array}{cc}
y & x \\ 
x & xz-y
\end{array}
\right)   \tag{2.9b}
\end{equation}
,etc. For the integer values of x, y and z the identity, Eq.(2.7), is known
as equation for the Markov triples discovered in the number theory by Markov
[36] . Given this equation, one is interested to obtain all integer
solutions (triples x, y and z). In mathematics literature, sometimes,
related equation is known as an equation for the Markov triples [39].
Specifically, one introduces the following redefinitions: 
\begin{equation*}
x=3m_{1},\text{ }y=3m_{2}\text{ and }z=3m_{3}
\end{equation*}
so that Eq.(2.7) acquires the standard form: 
\begin{equation}
m_{1}^{2}+m_{2}^{2}+m_{3}^{2}=3m_{1}m_{2}m_{3}.   \tag{2.10}
\end{equation}
The simplest solution of this equation is just $m_{1}=m_{2}=m_{3}=1.$To
generate other solutions it is sufficient to have a ''seed'' ($%
m_{1},m_{2},m_{3})$ which, by definition, obeys the Markov Eq.(2.10). Then,
the first generation of Markov triples is given by 
\begin{equation*}
(m_{1}^{\prime},m_{2},m_{3}),(m_{1},m_{2}^{\prime},m_{3})\text{ and (}%
m_{1},m_{2},m_{3}^{\prime}) 
\end{equation*}
where 
\begin{equation}
m_{1}^{\prime}=3m_{2}m_{3}-m_{1};m_{2}^{\prime}=3m_{1}m_{3}-m_{2};m_{3}^{%
\prime}=3m_{1}m_{2}-m_{3}.   \tag{2.11}
\end{equation}
Other solutions can be generated via mapping: ($m_{1},m_{2},m_{3})%
\rightarrow(-m_{1},-m_{2},m_{3})$ along with cyclical permutation of $%
m_{1},m_{2}$ and $m_{3}.$ If one starts with (1,1,1) then, the Markov tree
(its part, of course) is depicted in Fig.2.\FRAME{ftbpFU}{5.0721in}{5.8963in%
}{0pt}{\Qcb{Fragment of Markov tree close to the original ''seed''}}{}{fig.
2.gif}{\special{ language "Scientific Word"; type "GRAPHIC";
maintain-aspect-ratio TRUE; display "USEDEF"; valid_file "F"; width
5.0721in; height 5.8963in; depth 0pt; original-width 5.604in;
original-height 6.5207in; cropleft "0"; croptop "1"; cropright "1";
cropbottom "0"; filename 'New Folder/fig. 2.gif';file-properties
"XNPEU";}}

Emergence of numbers $m_{1}^{\prime}$, etc., can be easily understood if,
instead of Eq.(2.10), we would consider the following quadratic form: 
\begin{equation}
f(x)=x^{2}-3m_{2}m_{3}x+m_{2}^{2}+m_{3}^{2}   \tag{2.12}
\end{equation}
Then, for $f(x)=0$ we can identify $x=m_{1}$ and $x^{\prime}=m_{1}^{\prime
}=3m_{2}m_{3}-m_{1}.$ In physics literature the dynamics of Markov triples
has actually been studied already in connection with quasicrystals [24], 1d
tight binding Schr\"{o}dinger equations (with quasiperiodic potential),
etc.[40], and is known as dynamics of trace maps. This dynamics can be
easily understood based on Eq.s(2.10)-(2.12). In short, one studies maps $%
\mathcal{F}$ of the type 
\begin{equation}
\mathcal{F}\text{ : }\left( 
\begin{array}{c}
x \\ 
y \\ 
z
\end{array}
\right) \rightarrow\left( 
\begin{array}{c}
3yz-x \\ 
y \\ 
z
\end{array}
\right)   \tag{2.13}
\end{equation}
etc. which possess an integral of motion 
\begin{equation}
I(x,y,z)=x^{2}+y^{2}+z^{2}-3xyz   \tag{2.14}
\end{equation}
invariant under action of $\mathcal{F}$.

From the theory of Teichm\"{u}ller spaces [41] it is known, that the length $%
l(\gamma)$ of \textbf{closed} geodesics associated with $\gamma\in G$ is
given by 
\begin{equation}
tr^{2}(\gamma)=4\cosh^{2}\left( \frac{l(\gamma)}{2}\right) .   \tag{2.15}
\end{equation}
Let, $tr\gamma=x$ $($or $y,$ or $z)$, then the Markov triple (1,1,1)
corresponds to the geodesic whose hyperbolic length $l(\gamma)$ is given by 
\begin{equation}
l(\gamma)=2\cosh^{-1}(\frac{3}{2})=2\ln(\frac{1}{2}(3+\sqrt{5}))=2\ln
\lambda   \tag{2.16}
\end{equation}
where in the last equality use had been made of Eq(2.3). Obtained result
provides us with the first piece of evidence that the train tracks are
directly associated with closed geodesic laminations. Since in the case of
the punctured torus the Teichm\"{u}ller space coincides with the Poincar$%
e^{\prime }$ upper half plane model of hyperbolic space, the Teichm\"{u}ller
distance $d_{T}$ is the same as hyperbolic distance $l_{H}$ $[41].$
Therefore, we obtain, 
\begin{equation}
d_{T}=\frac{1}{2}\ln K=l_{H}=2\ln\lambda   \tag{2.17}
\end{equation}
which leads to equation 
\begin{equation}
\sqrt{K}=\lambda^{2}.   \tag{2.18}
\end{equation}
That is the stretching factor $\lambda$ is associated directly with the
Teichm\"{u}ller dilatation factor $K$ (for more details on $K$, please,
consult our earlier work, Ref.[17]$).$

In the light of previous discussion we notice that the spectrum of
stretching factors is discrete. This is reminiscent already to the energy
spectrum of some quantum mechanical system. The question arises at this
point: is the stretching factor $\lambda$, defined by Eq.(2.3), represents
the maximum or the minimum among possible stretching factors? A simple
calculation based on Eq.(2.15) and Fig.2 indicates that $\lambda,$ given by
Eq.(2.15), corresponds to the minimum. In the case of higher genus Riemann
surfaces (perhaps with punctures) Penner [42] had demonstrated that the
stretching factors are bounded from above and from below.

Let us now obtain the explicit form of matrices $a$ and $b$, given by
Eq.(2.9), for the case when $x=y=z=3$. An easy calculation produces: 
\begin{equation}
a=\left( 
\begin{array}{cc}
2 & 1 \\ 
1 & 1
\end{array}
\right) ,\text{ }b=\left( 
\begin{array}{cc}
2 & -1 \\ 
-1 & 1
\end{array}
\right) ,   \tag{2.19}
\end{equation}
yielding the trace of commutator $[a,b]$ being equal to -2 as required.
Introduce now two basic matrices 
\begin{equation}
L=\left( 
\begin{array}{cc}
1 & 0 \\ 
1 & 1
\end{array}
\right) \text{ and }R=\left( 
\begin{array}{cc}
1 & 1 \\ 
0 & 1
\end{array}
\right)   \tag{2.20}
\end{equation}
. The letters stand for the ''left'' ($L$) and the ''right'' ($R$) Dehn
twist matrices (e.g. see section 3 for more details). Evidently, 
\begin{equation}
a=RL,   \tag{2.21}
\end{equation}
as can be seen by direct calculation. It is also not difficult to check that 
\begin{equation}
b=L^{-1}aR^{-1}=L^{-1}RLR^{-1}.   \tag{2.22}
\end{equation}
Let us now take a note of the fact that 
\begin{equation}
L^{-1}=\left( 
\begin{array}{cc}
1 & 0 \\ 
-1 & 1
\end{array}
\right) \text{ and }R^{-1}=\left( 
\begin{array}{cc}
1 & -1 \\ 
0 & 1
\end{array}
\right) .   \tag{2.23}
\end{equation}
If we introduce the matrix 
\begin{equation}
\hat{I}=\pm\left( 
\begin{array}{cc}
0 & -1 \\ 
1 & 0
\end{array}
\right)   \tag{2.24}
\end{equation}
such that 
\begin{equation}
\hat{I}^{2}=I=\left( 
\begin{array}{cc}
1 & 0 \\ 
0 & 1
\end{array}
\right)   \tag{2.25}
\end{equation}
,then, Eqs.(2.23), (2.24) can be conveniently rewritten as 
\begin{equation}
L^{-1}=\hat{I}R\hat{I}   \tag{2.26}
\end{equation}
and 
\begin{equation}
R^{-1}=\hat{I}L\hat{I}\text{ }   \tag{2.27}
\end{equation}
which also can be checked by direct calculation. Thus obtained relations can
be equivalently presented as follows: 
\begin{equation}
L\hat{I}L=R,\text{ }R\hat{I}R=L,\text{ }R\hat{I}L=\hat{I}\text{ and }L\hat{I}%
R\text{=}\hat{I}\text{.}   \tag{2.28}
\end{equation}
Using the results just obtained, it is clear, that every word $W$ , e.g. see
Eq.(2,5), can be written in terms of positive powers of $L,R$ and $\hat{I}$.
Moreover, since originally we had only matrices $a$ and $b,$ things can be
simplified further. In particular, let us consider the word 
\begin{equation}
W_{1}=L^{\alpha_{1}}\cdot\cdot\cdot RL\cdot\cdot\cdot
L^{\alpha_{r}}R^{\beta_{r}}   \tag{2.29}
\end{equation}
and the related word 
\begin{equation}
W_{2}=L^{\alpha_{1}}\cdot\cdot\cdot R\hat{I}L\cdot\cdot\cdot
L^{\alpha_{r}}R^{\beta_{r}}.   \tag{2.30}
\end{equation}
We can eliminate the contribution $R\hat{I}L$ \ in Eq.(2.30) by replacing it
with $\hat{I}$ using Eq.s(2.28). If we continue to use Eq.s(2.28) we can
evidently get rid of all $\hat{I}$ factors in the middle of the word $W_{2}$
so that in the end we have to consider only the totality of words of the
type $W_{1}$with non$\func{negative}\func{integer}$ coefficients.To the
totality of these words one has to add words like $\hat{I}W_{1},W_{1}\hat{I}$
and $\hat{I}W_{1}\hat{I}$ . Since in section 4 we shall be interested in the
traces of these words, evidently, only words of the type $W_{1}$ and $\hat{I}%
W_{1}$ need to be considered. Incidentally, use of Eq.s(2.28) allows us to
reduce $b$, Eq.(2.22), to $b=\hat{I}LR\hat{I}$ and, hence, from now on we
shall use $a=RL$ and $b=LR$ . Such choice is in accord with that known in
the literature [43,44]. Moreover, to make connections with the results from
knot theory and 3-manifolds (to be discussed in sections 3 and 5) only words
of the type $W_{1}$ should be considered. This peculiarity will be explained
further below.

\subsection{\protect\bigskip The Farey numbers and the Farey tesselation of H%
$^{2}$}

Words of the type $W_{1}$ are represented by the set of matrices $M$, 
\begin{equation}
M=\left( 
\begin{array}{cc}
\alpha & \beta \\ 
\gamma & \delta
\end{array}
\right) ,\text{ }\alpha\delta-\gamma\beta=\pm1,   \tag{2.31}
\end{equation}
with integer coefficients. Such matrices belong to the group GL(2, Z) and
play an important role in the number theory [45]$.$ In the number theory
they are associated with the Farey numbers.\ Recall, that the Farey series
of numbers $\frak{F}_{n}$ is the ascending series of irreducible fractions
between 0 and 1 whose denominators do not exceed n. If p/q and r/s are two
consecutive terms (\textbf{neighbors}) of $\frak{F}_{n}$ , then 
\begin{equation}
ps-qr=\pm1.   \tag{2.32}
\end{equation}
In addition, if p/q , r/s and h/k are three consecutive terms, then the 
\textbf{mediant }r/s is obtained via 
\begin{equation}
\frac{r}{s}=\frac{p\pm h}{q\pm k}.   \tag{2.33}
\end{equation}
Hence, $\frac{\alpha}{\gamma}$ and $\frac{\beta}{\delta}$ in the matrix $M$
correspond to the neighbors in $\frak{F}_{n}.$ Given this information, one
can make two additional steps. First, one can let n go to $\pm\infty$ [46]$.$%
Thus, 
\begin{equation*}
\frak{F}_{1}:-\infty,-1,0,1,\infty\text{ ,}
\end{equation*}
\begin{equation*}
\frak{F}_{2}:-\infty,-2,-1,-\frac{1}{2},0,\frac{1}{2},1,2,\infty\text{ \ ,
etc.}
\end{equation*}
Second, by noticing that projectivisation PSL(2,R) of SL(2,R) is just an
isometry of H$^{2}$ it is advantageous to map conformally (-$\infty,\infty)$
into $S_{\infty}^{1}.$ Once this is done, locations of the Farey numbers on $%
S_{\infty}^{1}$ acquire new geometrical meaning (see also section 5.1). To
understand its significance, let us consider the M\"{o}bius-like
transformations associated with matrices $L$ and $R$, i.e. 
\begin{equation}
L:\text{ }z^{\prime}=\frac{z}{z+1}\text{ \ and }R\text{: }z^{\prime }=z+1.  
\tag{2.34}
\end{equation}
Let us notice first that for L-type transformation 
\begin{equation}
z=0\rightarrow z^{\prime}=0,\text{ }z=\pm\infty\rightarrow z^{\prime}=1\text{
and }z=-1\rightarrow z^{\prime}=-\infty   \tag{2.35a}
\end{equation}
At the same time, for R-type transformation we have 
\begin{equation}
z=0\rightarrow z^{\prime}=1,\text{ }z=\pm\infty\rightarrow
z^{\prime}=\pm\infty.   \tag{2.35b}
\end{equation}
Using these simple results we obtain at once 
\begin{equation}
b(\infty)=LR(\infty)=1,a(0)=RL(0)=1,a(-1)=RL(-1)=-\infty ,b(-1)=LR(-1)=0  
\tag{2.36}
\end{equation}
It is instructive to depict these results graphically, e.g.see Fig.3.\ 
\FRAME{ftbpFU}{2.437in}{2.4033in}{0pt}{\Qcb{Leaky torus in the Poicare disc $%
\mathcal{D}$ model.}}{}{fig.3.gif}{\special{ language "Scientific Word";
type "GRAPHIC"; maintain-aspect-ratio TRUE; display "USEDEF"; valid_file
"F"; width 2.437in; height 2.4033in; depth 0pt; original-width 5.354in;
original-height 5.2814in; cropleft "0"; croptop "1"; cropright "1";
cropbottom "0"; filename 'New Folder/fig.3.gif';file-properties
"XNPEU";}}\ 

\ \ \ \ \ \ \ \ \ \ \ \ \ \ \ \ \ \ \ \ \ \ \ \ \ \ \ \ \ \ \ \ \ \ \ \ \ \
\ \ \ \ \ \ \ \ \ \ \ \ \ \ \ \ \ \ \ \ \ \ \ \ \ \ \ \ \ \ \ \ \ \ \ \ \ \
\ \ \ \ \ \ \ \ \ \ \ \ \ \ \ \ \ \ \ \ \ \ \ \ \ \ \ \ \ \ \ \ \ \ \ \ \ \
\ \ \ \ \ \ \ \ \ \ \ \ \ \ \ \ \ \ \ \ \ \ \ \ \ \ \ \ \ \ \ \ \ \ \ \ \ \
\ \ \ \ \ \ \ \ \ \ \ \ \ \ \ \ \ \ \ \ \ \ \ \ \ \ \ \ \ \ \ \ \ \ \ \ \ \
\ \ \ \ \ \ \ \ \ \ \ \ \ \ \ \ \ \ \ \ \ \ \ \ \ \ \ \ \ \ \ 

Points -$\infty,-1,0,1,+\infty$ on the circle $S_{\infty}^{1}$ are just the
members of $\frak{F}_{1}.$These are joined by the circular arcs (not to be
mistaken for the hyperbolic geodesics which are going to be discussed
below). The arrows are in accord with the results given by Eq.s(2.36).
Identifying sides of the polygon using the arrows depicted in Fig.3 we
obtain an orbifold known in physics literature as ''leaky torus'' [13]. To
construct the next level of Farey numbers, $\frak{F}_{2},$and, for this
matter, $\frak{F}_{n},etc.,$ let us consider the most general modular
transformation 
\begin{equation}
z^{\prime}=\frac{az+b}{cz+d}\equiv M(z)   \tag{2.37}
\end{equation}
where the integer coefficients $a,b,c,d$ are subject to the constraint $\
ad-bc=1$(to be compared with Eq.s(2.31) and (2.32)).Using this
transformation we obtain, 
\begin{equation}
z=\pm\infty\rightarrow z^{\prime}=\frac{a}{c},\text{ }z=0\rightarrow
z^{\prime}=\frac{b}{d}\text{ and }z=\pm1\rightarrow z^{\prime}=\frac{\pm a+b%
}{\pm c+d}.   \tag{2.38}
\end{equation}
But the numbers $\frac{a}{c}$ and $\frac{b}{d}$ are just the Farey neigbors!
Using Eq.(2.33) we obtain the following sequence of numbers: 
\begin{equation}
-2=\frac{-1-1}{0+1},-\frac{1}{2}=\frac{-1+0}{1+1},\frac{1}{2}=\frac{0+1}{1+1}%
\text{ and 2=}\frac{1+1}{0+1}.   \tag{2.39}
\end{equation}
If we place these numbers on the circle $S_{\infty}^{1}$ we obtain all the
members of $\frak{F}_{2},etc.$as it is depicted in Fig.4.\FRAME{ftbpFU}{%
2.8461in}{2.5477in}{0pt}{\Qcb{Farey tesselation of $\mathcal{D}$.}}{}{%
fig.4.gif}{\special{ language "Scientific Word"; type "GRAPHIC";
maintain-aspect-ratio TRUE; display "USEDEF"; valid_file "F"; width
2.8461in; height 2.5477in; depth 0pt; original-width 5.2191in;
original-height 4.6665in; cropleft "0"; croptop "1"; cropright "1";
cropbottom "0"; filename 'New Folder/fig.4.gif';file-properties
"XNPEU";}}The question arises now: what combinations of $L$ and $R$ will
lead us to these numbers? Let us consider the generic case of $\frac{1}{2}.$
With help of Eq.s (2.38) and (2.39) we obtain two transformations 
\begin{equation}
z^{\prime}=\frac{z}{z+1}\text{ and }z^{\prime}=\frac{az+1}{cz+2}\text{ , }%
2a-c=1   \tag{2.40}
\end{equation}
connecting respectively the numbers 1 and 0 with the Farey number $\frac{1}{2%
}.$ Since both $a$ and $b$ are integers, the simplest choice is to take $%
a=1=c$.This then produces immediately: 
\begin{equation}
z^{\prime}=L(z)\text{ and }z^{\prime}=LR(z).   \tag{2.41}
\end{equation}
Other examples can now be constructed without problems. Evidently, each
successive level of $\frak{F}_{n}$ can be obtained by some application of
combinations of $L$'s and $R$'s. To make this procedure systematic, let us
consider some Farey number, say $p/q$, which has a continued fraction
expansion of the type [45] 
\begin{equation}
\frac{p}{q}=a_{0}+\frac{1}{a_{1}+\dfrac{1}{a_{2}+\dfrac{1}{a_{3}+\cdot
\cdot\cdot}}}\equiv\lbrack a_{0},a_{1},...,a_{n}]   \tag{2.42}
\end{equation}
where $a_{1},...,a_{n}$ are some integers. Suppose, we would like to connect
the point 0 (or $\infty)$ with the point $p/q.$ To this purpose, using
results given by Eq.(2.23), we need to take into account that 
\begin{equation}
x^{\prime}=L^{-1}x=\frac{x}{-x+1}=\frac{1}{\dfrac{1}{x}-1}\text{ and }%
x^{\prime}=R^{-1}x=x-1.   \tag{2.43}
\end{equation}
Using these results, it is clear, that 
\begin{equation}
R^{-a_{0}}(\frac{p}{q})=\frac{1}{a_{1}+\dfrac{1}{b_{1}}},   \tag{2.44}
\end{equation}
where 
\begin{equation}
\frac{1}{b_{1}}=\frac{1}{a_{2}+\dfrac{1}{c_{1}}},\text{ etc.}   \tag{2.45}
\end{equation}
Using Eq.s(2.43) and (2.44) we obtain, 
\begin{equation}
L^{-1}R^{-a_{0}}(\frac{p}{q})=\dfrac{1}{a_{1}+\dfrac{1}{b_{1}}-1}  
\tag{2.46}
\end{equation}
and, since 
\begin{equation}
L^{-a}x=\frac{1}{\dfrac{1}{x}-a}   \tag{2.47}
\end{equation}
we may as well write, instead of Eq.(2.46), 
\begin{equation}
L^{-a_{1}}R^{-a_{0}}(\frac{p}{q})=\frac{1}{b_{1}}.   \tag{2.48}
\end{equation}
Taking into account Eq.(2.45) we also get 
\begin{equation}
L^{-a_{2}}\left( L^{-a_{1}}R^{-a_{0}}(\frac{p}{q})\right) =a_{3}+\frac
{1}{d_{1}}.   \tag{2.49}
\end{equation}
Now one has to apply to this $R^{-a_{3}}$ in order to obtain expression
similar to Eq.(2.44) so that this process may continue. The end result will
depend upon wether $a_{n}>1$ or $a_{n}=1.$ When $a_{n}>1$ we can write $%
a_{n}=(a_{n}-1)+1$ and then use the operator $\ L^{-(a_{n}-1)}$ to $\dfrac
{1}{a_{n}}.$ Based on the results just obtained, it should be clear by now
that it is possible to construct any Farey number starting from 0,1,-1 or $%
\pm\infty$ with the result being some word W of the type 
\begin{equation}
W=L^{\alpha_{1}}R^{\beta_{1}}\cdot\cdot\cdot L^{\alpha_{r}}R^{\beta_{r}}  
\tag{2.50}
\end{equation}
where the exponents $\alpha_{1},\beta_{1},...,\alpha_{r},\beta_{r}$ are
related directly to numbers $a_{1},...,a_{n}.$

Consistent triangulation of H$^{2}(or$ $\mathcal{D}$ ) invariant with
respect to the action of PSL(2,Z) can be achieved now. To this purpose
first, we connect the point $\pm\infty$ with the point $\frac{0}{1}$ of
Fig.4 with help of the equation 
\begin{equation}
LRL^{-1}R^{-1}(-\infty)=0   \tag{2.51}
\end{equation}
which follows easily from Eq.(2.36). Since the points $\frac{1}{1}$ and $%
\frac{-1}{1}$ are mediants already, it should be clear how to produce the
rest of the triangles from these two which are basic as it is depicted in
Fig.4. By direct observation, it is easy to notice that this figure has an
axial symmetry. It can be shown [44], that the r.h.s corresponds to words of
type W$_{1}$ while the l.h.s. corresponds to words of the type $\hat{I}W_{1}$
and that these two sets are nonoverlapping. Hence, we shall use here and in
sections 3 and 4 only the r.h.s.of this figure. Although lines on this
figure look like geodesics, actually, they are not geodesics in a usual
sense (so that the statement made in Ref.[47], e.g. see \ page 566\ , that
these lines are geodesics should be treated with caution ) as it will be
explained below. Hence, the triangulation in Fig.4 should be understood only
in a topological sense (to keep track of how the Farey numbers are related
to each other). An alternative and very effective geometrical description of
Farey numbers (via the circle packing in H$^{2})$ had been proposed by
Rademacher [48 ]. We shall discuss it \ briefly in section 5.

Although the Farey tesselation of $\mathcal{D}$ depicted in Fig.4 is
helpful, it cannot be used directly for our calculations since we are
interested in geodesics related to the Markov triples. Therefore now we
would like to relate these Markov geodesics to the Farey tesselation. The
simplest Markov geodesics are associated with matrices given by Eq.(2.19)
(in view if Eq.(2.15)). Since we require these matrices to correspond to the 
\textbf{closed} geodesics on the Riemann surface of leaky torus, when lifted
to $\mathcal{D}$ (or H$^{2}),$their ends must lie on $S_{\infty}^{1}$ (or on
the line y=0 in H$^{2}$ model ).The location of the ends of geodesics is
determined by the fixed points equation which in the case of matrix $a$ is
given by 
\begin{equation}
x=\frac{2x+1}{x+1}=1+\frac{1}{1+\dfrac{1}{x}}.   \tag{2.52}
\end{equation}
Iteration of this equation leads to the\textbf{\ periodic} continued
fraction expansion with period 1 [45] 
\begin{equation}
x=[\dot{1}]=\frac{\sqrt{5}+1}{2}.   \tag{2.53}
\end{equation}
We had used the notations of Ref.[45], e.g. see. page 46, to reflect the
periodicity. Surely, this result can be obtained as well directly from the
quadratic equation 
\begin{equation}
x^{2}-x-1=0   \tag{2.54}
\end{equation}
which is equivalent to Eq.(2.52). Eq.(2.54) has two roots 
\begin{equation}
x_{1,2}=\frac{1}{2}(1\pm\sqrt{5}).   \tag{2.55}
\end{equation}
To recover the second root from Eq.(2.52) we have to introduce the following
change of variable: x=-$\dfrac{1}{y}$, in Eq.(2.52).This then produces after
a few trivial manipulations 
\begin{equation}
y=-\frac{1}{1+\dfrac{1}{1-y}}.   \tag{2.56}
\end{equation}
Iterating this expression we obtain, 
\begin{equation}
y=-\frac{1}{1+\dfrac{1}{1+\dfrac{1}{1+\cdot\cdot\cdot}}}.   \tag{2.57}
\end{equation}
This is equivalent to the continued fraction expansion of $\frac{1}{2}(1-%
\sqrt{5})$ [45] as required. Obtained results are special cases of general
theorem proven by Series [46] which states that 
\begin{equation}
x_{1}=x_{\infty}=[n_{1},n_{2},...]\text{ and }x_{2}=x_{-\infty}=\frac
{-1}{[n_{0},n_{-1},n_{-2},...]}   \tag{2.58}
\end{equation}
so that in \textbf{both} cases we have an \textbf{infinite} continued
fractions corresponding to irrational numbers. In the case of fixed points
for the Markov matrices the continued fractions are \textbf{always} periodic
[49].This fact has profound topological significance as it will be explained
in section 3. In the meantime, we still need to clarify the connections
between the Farey tesselation of H$^{2}$(or $\mathcal{D}$) and the results
just obtained. The comprehensive treatment of this problem by
mathematicians, surprisingly, had been performed only quite recently
[43,47]. Moreover, to our knowledge, there are no similar comprehensive
treatments for surfaces of higher genus (perhaps, with exception of
Ref.[50]). The major reason for utilization of the Farey triangulation of H$%
^{2}$ is exactly the same as used in approximations of irrational numbers by
the rationals which belong to the Farey series [45]. According to the theory
of numbers [45] \textbf{all} rational numbers are \textbf{equivalent }since
they are connected with\textbf{\ }each\textbf{\ }other via modular
transformation, Eq.(2.37).\textbf{\ }In our case this means that the
vertices of the basic quadrangle depicted in Fig.3 upon identification
(needed for formation of the leaky torus) all correspond to a puncture.
Subsequent higher levels of the Farey tesselations do not change this
result: all Farey numbers correspond to a puncture. Hence, from here it also
follows that the ends of geodesics which correspond to the \textbf{%
nonperipheral} elements of $G$ are represented by the irrational numbers
connected by bi infinite sequence of M\"{o}bius transformations. In
particular case of Eq.(2.52) this becomes obvious if we rewrite it in the
form 
\begin{equation}
x=RL(x).   \tag{2.59}
\end{equation}
Iterating, we obtain, 
\begin{equation}
x=RLRLRL\cdot\cdot\cdot(x).   \tag{2.60}
\end{equation}
To clarify the meaning of this result, please, recall that , according to
Eq.(2.21), $RL=a$.The eigenvalues $\lambda_{1,2}$ of $a$ can be easily found
from the equation 
\begin{equation}
\lambda^{2}-3\lambda+1=0   \tag{2.61}
\end{equation}
thus producing 
\begin{equation}
\lambda_{1,2}=\frac{1}{2}(3\pm\sqrt{5}).   \tag{2.62}
\end{equation}
Not surprisingly, $\lambda_{1}$coincides with earlier obtained result,
Eq.(2.3). These results will acquire completely new topological meaning in
section 3.

Consider now the M\"{o}bius transformation $U_{\lambda}(z)$ given by 
\begin{equation}
z^{\prime}=U_{\lambda}(z)=\left( \frac{\lambda_{1}}{\lambda_{2}}\right)
z\equiv\hat{\lambda}z.   \tag{2.63}
\end{equation}
This transformation has two fixed points : z$^{\ast}=0$ and $z^{\ast}=\infty.
$ Consider yet another M\"{o}bius transformation $W(z)$ such that 
\begin{equation}
W(x_{1})=0\text{ and }W\text{(}x_{2}\text{)}=\infty,\text{ e.g. }W(z)=\frac{%
z-x_{1}}{z-x_{2}},   \tag{2.64}
\end{equation}
where $x_{1}$ and $x_{2}$ are given by Eq.(2.55). Hence, using such
transformation, we obtain, 
\begin{equation}
WaW^{-1}(z)=U_{\lambda}(z),   \tag{2.65}
\end{equation}
so that Eq.(2.60) can be rewritten as 
\begin{equation}
x=WaW^{-1}WaW^{-1}\cdot\cdot\cdot(x)=U_{\lambda}U_{\lambda}\cdot\cdot\cdot
U_{\lambda}(x).   \tag{2.66}
\end{equation}
Therefore, indeed, in view of Eq.(2.63), we obtain the expected sequence of
iterates connecting two fixed points. These points are the initial and the
final limiting points representing ''motion'' along the geodesics in H$^{2}$
which are just semicircles passing through points $x_{1}$ and $x_{2}.$ It is
very important to realize that in the case of a torus (and also punctured
torus) the hyperbolic H$^{2}$ plane coincides with the Teichm\"{u}ller
space. Hence, the ''motion'' along the geodesic in H$^{2}$ coincides with
the \textbf{real} motion in the Teichm\"{u}ller space as was discussed
qualitatively in our earlier work on 2+1 gravity. More details will be
provided in sections 3 and 4.

Consider now the possibility of joining of two Farey numbers by the
geodesics. That is, let us consider the fixed point of equation 
\begin{equation}
x=\frac{ax+b}{cx+d},\text{ }ad-cb=1.   \tag{2.67}
\end{equation}
We need to look for solution of this equation only for the \textbf{integer}
values of $a$, $b$, $c$ and $d$.\ It is instructive to discuss some special
cases first. Let us begin with the case $c=b=0.$ In this case we obtain, 
\begin{equation}
x=\frac{a}{d}x.   \tag{2.68}
\end{equation}
Although this equation looks exactly like Eq.(2.63), these equations are 
\textbf{not} the same since according to Eq.(2.67) we should require $ad=1$
which leaves us with the only choice: $a=d=\mp1.$The obtained identity
result indicates that the hyperbolic-like transformations are \textbf{not}
possible if we use just PSL(2,Z). Let us now consider the parabolic
transformations, e.g. let $c=0$ then, given that $ad=1,$ we obtain, 
\begin{equation}
x=x+b.   \tag{2.69}
\end{equation}
This equation has only one fixed point, x$^{\ast}=\infty$ , characteristic
for all parabolic transformations. Let us now put $b=0$ in Eq.(2.67).Then,
again, $ad=1$ and we obtain, 
\begin{equation}
x=\frac{x}{cx+1},   \tag{2.70}
\end{equation}
thus producing another acceptable solution: x$^{\ast}=0.$ Finally, let $d=0$
or $a=0$. In the first case we obtain cb=-1 so that 
\begin{equation}
x=\frac{ax-1}{x}   \tag{2.71}
\end{equation}
thus leading to the equation 
\begin{equation}
x_{1,2}=\frac{a}{2}\pm\frac{1}{2}\sqrt{a^{2}-4},   \tag{2.72}
\end{equation}
while in the second case we get $bc=-1$ so that 
\begin{equation}
x=\frac{-1}{x+d}   \tag{2.73}
\end{equation}
thus producing 
\begin{equation}
x_{1,2}=-\frac{d}{2}\pm\frac{1}{2}\sqrt{d^{2}-4}.   \tag{2.74}
\end{equation}
In both cases the problem is reduced to finding the Pythagorean numbers,
i.e. to finding all integer solutions of equation 
\begin{equation}
a^{2}+b^{2}=c^{2}.   \tag{2.75}
\end{equation}
As known results indicate [51], the only solution for d in Eq.(2.74) is d=2
( accordingly, $a=2$ in Eq.(2.72)) thus leaving us with just \textbf{one }\
fixed point. Clearly, we can identify thus obtained fixed points with 0,
1,-1 and $\pm\infty$ as depicted in Fig.3 since all other points can be
obtained by successive applications of modular transformations. Hence, the
semicircles depicted in Fig.4 are \textbf{not} true geodesics contrary to
the statements made in Ref.[47]. Consideration of the general case (section
3) produces the same negative result: only one trivial solution of
Eq.(2.75), \ thus leading to just one (parabolic) fixed point for PSL(2,Z).

Although we had provided enough evidence which connects the geodesic
laminations and the train tracks, more systematic treatment of this subject
would lead us somewhat away from the topics which we had discussed so far.
Fortunately, sections 1.5-1.7 of Ch-r 1 of the book by Penner and Harer [53]
contain all the required proofs (please, read especially pages 87-101). Much
shorter proofs could be also found in the unpublished PhD thesis by Lok
[54]. Since these results are needed only for rigorous mathematical proofs
of connections between laminations and train tracks, we hope, that our
readers will consult these references for better understanding of the
obtained results and those which follow in the rest of this paper.

\section{From geodesic laminations to 3-manifolds which fiber over the circle%
}

\subsection{\protect\bigskip The Alexander polynomial and surface
homeomorphisms}

The Alexander polynomial $\Delta_{8}(t)$ for the figure eight knot is known
to be [54] 
\begin{equation}
\Delta_{8}(t)=t^{2}-3t+1   \tag{3.1}
\end{equation}
Consider, quite formally (for the time being only!), the zeroes of this
polynomial. These are obtained as roots of the equation 
\begin{equation}
t^{2}-3t+1=0.   \tag{3.2}
\end{equation}
A simple calculation produces: 
\begin{equation}
t_{1,2}=\frac{1}{2}(3\pm\sqrt{5}).   \tag{3.3}
\end{equation}
Eq.s (3.2) and (3.3) we would like now to compare with Eq.s(2.61) and (2.62)
respectively. Since this comparison yields complete coincidence of results
the rest of this section is devoted to the proof \ that the above
coincidence is not accidental. As a result of such proof the connection
between the dynamics of 2+1 gravity and 3-manifolds is naturally
established. Unlike Witten's treatment of 2+1 gravity [55] which establish
this connection through reformulation of this problem in terms of the
Chern-Simons field theory, our treatment does not require field-theoretic
arguments at all and is based mainly on Thurston's theory of 3-manifolds
[19]. The condensed summary of relevant results of Thurston [19] and
McMullen [56] is given in our earlier publications, Ref.[16,17].

We would like to begin with reviewing \ some facts about the Alexander
polynomial. Although our earlier published review [57] provides sufficient
physical background on knot polynomials and, in particular, on the Alexander
polynomial, this background is not sufficient for our current purposes.
Hence, we shall avoid references to physics literature when we shall talk
about the Alexander polynomial. To simplify matters, we shall treat only the
case of the Alexander polynomials for knots. The case of links is
considerably more complicated and will be treated in a separate publication.
For the case of knots, let $V$ be the Seifert matrix of linking coefficients
[18,58], then the Alexander polynomial $\Delta_{K}(t)$ for a knot $K$ is
given by 
\begin{equation}
\Delta_{K}(t)=\det(V^{T}-tV)   \tag{3.4}
\end{equation}
where $V^{T}$ is the matrix transpose of $V$. Such defined polynomial has
some additional remarkable properties [18,58]. For instance, 
\begin{equation}
\Delta_{K}(1)=\pm1   \tag{3.5}
\end{equation}
\begin{equation}
\Delta_{K}(t)\dot{=}\Delta_{K}(t^{-1})   \tag{3.6}
\end{equation}
where the symbol \.{=} denotes an equality up to a constant multiplier (e.g.
see Eq.(3.1) for an obvious example). The above properties of $\Delta_{K}$
allow us to write it as polynomial of even degree $2r$ with integer
coefficients $a_{i}$ : 
\begin{equation}
\Delta_{K}(t)=\sum\limits_{i=0}^{2r}a_{i}t^{i}\text{ , }a_{2r-i}=a_{i}\text{
.}   \tag{3.7}
\end{equation}
Using the inversion symmetry property \ reflected in Eq.(3.6) the following
Laurent expansion for $\Delta_{K}$ can be written: 
\begin{equation}
\Delta_{K}(t)\dot{=}a_{r}+a_{r+1}(t+t^{-1})+...+a_{2r}(t^{r}+t^{-r}).  
\tag{3.8}
\end{equation}
We are not interested in \textbf{all} possible knots and their Alexander
polynomials since \textbf{not} all knots are of relevance to surface
dynamics. As we had discussed earlier [17], only \textbf{fibered} knots are
of relevance. The easiest way to talk about fibered knots embedded in $S^{3}$
is trough consideration of their complements $S^{3}\setminus K$ in $S^{3}.$
These are just 3 manifolds fibering over the circle $S^{1}$. Indeed, let $S$
be the Seifert surface associated with knot $K.$ Incidentally, the punctured
torus is the Seifert surface for both the figure eight and the trefoil
(right and left) knots [18,58]. There is another knot, the bridge knot
b(7,3), which also has the same Seifert surface but it cannot be fibered
over the circle [59]. The Alexander polynomial $\Delta_{T}$ for the trefoil
is known to be [54] 
\begin{equation}
\Delta_{T}(t)=t^{2}-t+1.   \tag{3.9}
\end{equation}
The zeros of this polynomial are readily obtained: 
\begin{equation}
t_{1,2}=\frac{1}{2}(1\pm i\sqrt{3}).   \tag{3.10}
\end{equation}
Apparently, they have nothing to do with the discussion made in the previous
section.This, however, is not true as we shall soon demonstrate. To this
purpose consider an orientation preserving surface homeomorphism $h$: $%
S\rightarrow S.$ In the case of the punctured(holed) torus the homeomorphism
should respect the presence of a hole. The circumference of this hole is
just our base space $S^{1}($which, as it is not too difficult to guess, is
just our knot $K$ since the knot is just a circle embedded into $S^{3}).$
The Seifert surface itself is a fiber and the 3-manifold is just a fiber
bundle constructed in a following way. Begin with products $S\times0$ (the
initial state) and $S_{h}\times1$ (the final state) so that for each point $%
x\in S$ we have $(x,0)$ and $(h(x),1)$ respectively. The interval $I=(0,1)$
can be now closed (to form a circle $S^{1})$ by identifying 0 with 1 which
causes identification: 
\begin{equation}
(x,0)=(h(x),1).   \tag{3.11}
\end{equation}
The fiber bundle (also known in the literature as \textit{mapping torus}
[56,60]) 
\begin{equation}
T_{h}=\left( S\times I\right) /h   \tag{3.12}
\end{equation}
is the 3-manifold which fibers over the circle and is complementary to the
fibered knot in $S^{3}.$ The interval (0, 1) can be associated with some
local time. The cyclic character of the process leading to formation of
3-manifold(s) is not essential as it will be demonstrated below.Therefore,
actually, the time interval can be taken from -$\infty$ to $\infty$ . The
periodicity naturally occurs, if homeomorphisms are associated with motion
along the Markov geodesics in the Teichm\"{u}ller space, e.g. see Eq.(2.60).
Indeed, the homeomorphisms of surfaces are associated with dynamics of train
tracks. If we start with train track dynamics, it is in one-to one
correspondence with dynamics of geodesic laminations and this dynamics, in
turn, is associated with motions in Teichm\"{u}ller space, e.g. along some
geodesics in this space. Hence, the periodicity occurs quite naturally.
Othal [61] and McM\"{u}llen [56] had proved that the situation just
described for the punctured torus persist for Riemann surface (with marking
and /or boundaries) of any genus g as it was briefly mentioned earlier in
our work, Ref.[17]. More specifically, they proved the following theorem

\bigskip

\textbf{Theorem 3.1.} \textit{Let }$\psi:$\textit{\ }$S\rightarrow S$\textit{%
\ be a pseudo-Anosov homeomorphism of compact surface with negative Euler
characteristic. Then, in order for the mapping torus }$T_{\psi}$\textit{\ to
have a hyperbolic structure one is looking for the related hyperbolic
manifold }$M_{\psi}=S\times R$\textit{\ on which the homotopy class of }$\psi
$\textit{\ is represented by an isometry }$\alpha$\textit{\ . Then, }$%
M_{\psi}/<\alpha>$\textit{\ is homeomorphic to }$T_{\psi}.$

\bigskip

It can be demonstrated, e.g. Proposition 5.10 and the comments which follow
in Ref.[18], that classification of \textbf{all} fibered knot complements
can be formulated in terms of fibering surfaces and maps of such
surfaces.Therefore, naturally, this fact is reflected in the associated
Alexander polynomials. Unfortunately, the conjecture that all fibered knots
are classified one-to-one by their Alexander polynomials happens to be
wrong. Morton had demonstrated [62] that for the Seifert surfaces of genus g%
\TEXTsymbol{>}1 there are \textbf{infinitely many different fibered} \textbf{%
knots} for \textbf{each} Alexander polynomial of degree \TEXTsymbol{>}2.
This fact by no means diminishes the role of the Alexander polynomial in
dynamics of 2+1 gravity, it just makes knot interpretations [22,55] of
gravity less convincing.\ Similar negative conclusions had been reached
recently with help of absolutely different set of arguments in Ref.[63].

From the knot theory [18 ] it is known, that instead of expansion, Eq.(3.7),
for the Alexander polynomial for \textbf{any} knot, in the case of \textbf{%
fibered} knots one has to use 
\begin{equation}
\Delta_{K}(t)=\sum\limits_{i=0}^{2g}a_{i}t^{i},   \tag{3.13}
\end{equation}
where g is the genus of the associated Seifert surface. Moreover, 
\begin{equation}
\Delta_{K}(0)=a_{0}=a_{2g}=\pm1,   \tag{3.14}
\end{equation}
,i.e. the Alexander polynomial is monic.Using Eq.(3.4) we obtain at once 
\begin{equation}
\Delta_{K}(0)=\det(V^{T})=\det(V)=\pm1   \tag{3.15}
\end{equation}
to be compared with determinant in Eq.(2.31). Using Eqs.(3.15) and (3.4) we
obtain without delay, 
\begin{equation}
\Delta_{K}(t)=\det(V^{-1}V^{T}-tE)\equiv\det(M-tE),   \tag{3.16}
\end{equation}
where $E$ is the unit matrix and $M$ is the monodromy matrix responsible for
surface automorphisms [18, 58]. This fact easily follows from the \
observation \ that $\Delta_{K}(0)=\det(M)=\pm1.$

\bigskip

\textbf{Remark 3.2.}\ Equating the Alexander polynomial to zero and finding
the roots of this polynomial is equivalent to solving the eigenvalue problem
for surface automorphisms which produces the stretching factors allowing one
to distinguish between the hyperbolic, pseudo-Anosov-type ( when the largest
root is real and greater than one), and the periodic, Seifert fibered-type
(when the modulus of the largest root is equal to one), regimes of surface
homeomorphisms.

\bigskip

Let us discuss Eq.(3.16) a bit more. It can be easily shown [64] that, if \ $%
M\in GL(2,Z),$then 
\begin{equation}
\Delta_{K}(t)=t^{2}-\left( trM\right) t+\det M.   \tag{3.17}
\end{equation}
But, as we know already, $\det M=\pm1$ . Hence, in this case we obtain,
instead of Eq.(3.17), 
\begin{equation}
\Delta_{K}(1)=t^{2}-\left( trM\right) t\pm1.   \tag{3.18}
\end{equation}
Using Eq(3.5) we obtain as well 
\begin{equation}
\Delta_{K}(1)=1-\left( trM\right) \pm1=\pm1.   \tag{3.19}
\end{equation}
This leaves us with two options: $tr(M)=1$ or 3. In the first case we
reobtain the Alexander polynomial for the trefoil, Eq.(3.9), and in the
second for the figure eight, Eq.(3.1),\ knots. No other options are
available! Hence we had just proved (in a somewhat different way as compared
with Ref.[18 ]), that the trefoil and the figure eight knots are the only
two fibered knots associated with the Sefert surfaces of genus 1. Moreover,
the conditions $trM=3$ and $detM=1$ lead us directly to the Markov matrix $a$%
, Eq.(2.19). This matrix, Eq.(2.60), and the Theorem 3.1 provide direct
connection between the hyperbolic 3-manifold associated with complement of
figure eight knot and the pseudo-Anosov surface homeomorphisms associated
with $a$.These conclusions are in accord with results of Thurston [19], e.g
see section 4.37, where they had been obtained in a different way.

At the same time, the conditions $trM=1$ and $detM=1$ are also very
interesting since they are associated with the matrix 
\begin{equation}
M=\left( 
\begin{array}{cc}
1 & 1 \\ 
-1 & 0
\end{array}
\right) \text{ or }\left( 
\begin{array}{cc}
1 & -1 \\ 
1 & 0
\end{array}
\right) .   \tag{3.20}
\end{equation}
whose eigenvalues we had calculated already in Eq.(3.10). It is easy to
check that such surface homeomorphism is \textbf{not} associated with motion
along the hyperbolic geodesic since the fixed point equation 
\begin{equation}
x=1-\frac{1}{x}   \tag{3.21}
\end{equation}
which is equivalent to Eq.(3.9) (if we require $\Delta_{T}(t)$ to be zero)
does not have real roots.Using the same methods as in section 2 we easily
obtain that projectively the matrix $M$ is equivalent to the combination of $%
LR^{-1}.$ Such transformations do not fit the Theorem 3.1 and the associated
3 manifolds are known as Seifert fibered spaces. Since the trefoil knot is
the simplest representative of torus knots, it can be demonstrated [18],
that the complements of \textbf{all} torus knots in $S^{3}$ are associated
with the Seifert fibered spaces. According to Thurston's classification of
surface homeomorphisms those leading to the Seifert-fibered spaces are known
as \textbf{periodic. }The detailed structure of the periodic phase may be
very complicated [65-67] and, hence, very interesting from the point of view
of physical applications. The discussion of all emerging possibilities,
surely, requires separate publications. The geometrical richness of the
Seifert-fibered regime provides a likely explanation of different states of
order in the case of liquid crystals as we had briefly mentioned in the
Introduction and earlier works. \textbf{\ }E.g.solid and hexatic phases
might be associated with the Seifert-like while liquid and/or gas phases may
be associated with the pseudo-Anosov-like. In this paper we shall be
concerned only with transition between the pseudo-Anosov and the
Seifert-like phases leaving the detailed analysis of possibilities emerging
in the Seifert-fibered (periodic) phase for future publications. In order
for our results to be consistent with the rest of the literature on phase
transitions we need now to introduce several new concepts.

\subsection{ Mahler measures and topological entropies}

Following Ref.[68] \ let us consider a monic polynomial with integer
coefficients 
\begin{equation}
F(x)=x^{d}+a_{d-1}x^{d-1}+...+a_{1}x+a_{0}   \tag{3.22}
\end{equation}
where $a_{0}=\pm1.$ If $\alpha_{i}$ are the roots of this polynomial, then,
equivalently, we can rewrite it as 
\begin{equation}
F(x)=\prod\limits_{i=1}^{d}(x-\alpha_{i}).   \tag{3.23}
\end{equation}
The logarithmic Mahler measure $m(F)$ can be defined now as 
\begin{equation}
m(F)=\ln M(F)   \tag{3.24}
\end{equation}
where $M(F)$ is given by 
\begin{equation}
M(F)=\prod\limits_{i=1}^{d}\max\{1,\left| \alpha_{i}\right| \}.   \tag{3.25}
\end{equation}
The following theorem is attributed to Walters, Ref.[69], section 8.4.

\bigskip

\textbf{Theorem 3.3.} \textit{The topological entropy of the transformation
M is equal to the logarithmic Mahler measure of the characteristic
polynomial of the matrix M.}

\bigskip

To make a connection with statistical mechanics, we would like to notice
that the fact the polynomial $F(x)$ is monic is not at all restrictive (we
used it only in order to make connection with the previous discussion). Let
us recall at this point some facts from Lee and Yang theory of phase
transitions [7]. For instance, for a gas of $\frak{M}$ atoms which can be
packed into the volume $\frak{v}$ the grand canonical partition function $\Xi
$ is given in a usual manner as 
\begin{equation}
\Xi=\sum\limits_{n=0}^{\frak{M}}\frac{Q_{n}}{n!}z^{n}.   \tag{3.26}
\end{equation}
Hence, the grand partition function is just some polynomial in fugacity $z.$
Surely, the polynomial, Eq.(3.26), must have some zeros. These zeros can be
only in the complex plane $\mathbf{C}$ and they have to come in pairs of
complex conjugates for Eq.(3.26) to be real for real z's. Hence,
effectively, we can rewrite $\Xi$ as follows 
\begin{equation}
\Xi(z)=\frac{Q_{\frak{M}}}{\frak{M!}}\prod\limits_{n=1}^{\frak{M/2}}(z-\bar
{z}_{n})(z-z_{n}).   \tag{3.27}
\end{equation}
Since $\Xi(z=0)=1,$the above result can be conveniently rewritten as 
\begin{equation}
\Xi(z)=\prod\limits_{n=1}^{\frak{M}}(1-\frac{z}{z_{n}})(1-\frac{z}{\bar{z}%
_{n}}).   \tag{3.28}
\end{equation}
For physical applications one is interested in finding of $ln\Xi$ that is 
\begin{equation}
\frak{F(}z)\frak{=}\ln\Xi\frak{=}\sum\limits_{n=1}^{\frak{M/2}%
}\ln(z^{2}-2\cos\theta_{n}+1)   \tag{3.28}
\end{equation}
where use had been made of the fact that the complex zeros should lie on the
unit circle in the complex plane \textbf{C }(this is demonstrated below). If
g($\theta)$ is the density of these zeros, one can replace the above
summation in Eq.(3.28) by the integration with the result 
\begin{equation}
\frak{F(}z)\frak{=}\int\limits_{0}^{\pi}d\theta g(\theta)\ln\left|
(z^{2}-2\cos\theta+1)\right|   \tag{3.29}
\end{equation}
which is written with account of symmetry of the integrand. To make a
connection between this result and the logarithmic Mahler measure it is only
sufficient to use

\bigskip

\textbf{Lemma 3.4 }(Mahler's lemma)[68] \textit{For any nonzero} $F\in C[x],$%
\begin{equation}
m(F)=\int\limits_{0}^{1}d\theta\ln\left| F(\exp(2\pi i\theta))\right| .  
\tag{3.30}
\end{equation}

\textbf{Proof}. This result follows at once if one can prove Jensen's
formula: for any $\alpha\in\mathbf{C}$ 
\begin{equation}
\int\limits_{0}^{1}\ln\left| \exp\left( 2\pi i\theta\right) -\alpha\right|
=\ln\max\{1,\left| \alpha\right| \}.   \tag{3.31}
\end{equation}
The detailed proof of this result can be found in Ref.[68].Using
Eq.s(3.23)-(3.25) the Lemma 3.4. is proved.

From here, the following theorem follows at once.

\bigskip

\textbf{Theorem 3.5}. $m(F)=0$ \textit{if and only if }$\left| \alpha
_{i}\right| =1$\textit{\ for all }$i$\textit{\ in Eq.(3.23), provided that }$%
F(x)$\textit{\ is monic polynomial. }

\textit{\bigskip}

This result coincides with that of Lee and Yang, e.g.see Theorem 3, e.g. see
page 418, Eq.(57) of Ref.[7], where it was obtained using different methods.
Theorem 3.5. does not imply that for \textbf{all} monic polynomials $m(F)=0$%
.This is definitely not the case for the Alexander polynomial, Eq.(3.1), for
the figure eight knot. Evidently, for such knot the topological entropy is
given by 
\begin{equation}
m_{8}=\ln\frac{1}{2}(3+\sqrt{5})   \tag{3.32}
\end{equation}
while for the trefoil knot $m_{T}$, is, indeed, zero in view of Eq.(3.10).
Hence, \textbf{the logarithmic Mahler measure can be used for the Alexander
polynomial of any fibered knot to distinguish between the pseudo-Anosov and
the periodic regimes.}We would like now to complicate matters by considering

\subsection{The incompressible surfaces in the once punctured torus bundles
over S$^{1}$}

Although the previous analysis is interesting in its own right, the
situations which we had considered so far (with the trefoil and the figure
eight knots) are not the only possibilities for the punctured torus
automorphims. To go beyond these possibilities means to abandon almost
completely the connections with knots and to concentrate attention on
solutions of the equation 
\begin{equation}
\frak{D}\text{ }(t)=\det(M-tE)=0.   \tag{3.33}
\end{equation}
Since we now do not require $\frak{D}$ (t) to be connected with knot
polynomials, there is no need for $\frak{D}$ $(1)$=$\pm1.$ At the same time,
Eq.(3.33) is legitimate equation for finding of stretching factors
characteristic for pseudo-Anosov homeomorphisms. For an arbitrary matrix $%
M\in SL(2,Z)$ we obtain, instead of Eq.(3.18), the equation ( $n\in Z$ ) 
\begin{equation}
t^{2}-nt\pm1=0   \tag{3.34}
\end{equation}
which produces the following roots: 
\begin{equation}
t_{1,2}=\frac{1}{2}(n\pm\sqrt{n^{2}\mp4}).   \tag{3.35}
\end{equation}
From here, we see that only n which obey $\left| n\right| \leq2$ produce the
stretching factors characteristic for the periodic (Seifert fibered) phase.
For all $\left| n\right| $ $\geq3$ the stretching factors are characteristic
of the pseudo-Anosov phase. These results can be interpreted topologically
with help of the notion of incompressible branched surfaces introduced by
Oertel [70]. We would like to discuss these surfaces in order to make
connections with the results of section 2 and in order to provide
topological interpretation of the results of section 4.

The notion of branched surfaces is inseparably connected with the notion of
train tracks. As soon as we would like to visualize the dynamics of train
tracks in time, we encounter branched surfaces which are perpendicular to
the original surface as it is depicted in Fig.5.\FRAME{ftbpFU}{4.3829in}{%
2.2943in}{0pt}{\Qcb{Branched surfaces associated with dynamics of train
tracks}}{}{fig.5.gif}{\special{ language "Scientific Word"; type "GRAPHIC";
maintain-aspect-ratio TRUE; display "USEDEF"; valid_file "F"; width
4.3829in; height 2.2943in; depth 0pt; original-width 12.2501in;
original-height 6.3754in; cropleft "0"; croptop "1"; cropright "1";
cropbottom "0"; filename 'New Folder/fig.5.gif';file-properties
"XNPEU";}}\ \ \ \ \ \ \ \ \ \ \ \ \ \ \ \ \ \ \ \ \ \ \ \ \ \ \ \ \ \ \ \ \
\ \ \ \ \ \ \ \ \ \ \ \ \ \ \ \ \ \ \ \ \ \ \ \ \ \ \ \ \ \ \ \ \ \ \ \ \ \
\ \ \ \ \ \ \ \ \ \ \ \ \ \ \ \ \ \ \ \ \ \ \ \ \ \ \ \ \ \ \ \ \ \ \ \ \ \
\ \ 

Naturally, these branched surfaces inherit the weights, e.g. see Fig.1, from
the associated with them train tracks and these weights obey the switch
conditions as for the train tracks. This is depicted in Fig.6a). \FRAME{%
ftbpFU}{5.0635in}{2.2269in}{0pt}{\Qcb{a) Kirkhoff -type rules for branched
surfaces . b) Another interpretation of the weights on branches (read the
text for details)}}{}{fig.6.gif}{\special{ language "Scientific Word"; type
"GRAPHIC"; maintain-aspect-ratio TRUE; display "USEDEF"; valid_file "F";
width 5.0635in; height 2.2269in; depth 0pt; original-width 12.3331in;
original-height 5.3852in; cropleft "0"; croptop "1"; cropright "1";
cropbottom "0"; filename 'New Folder/fig.6.gif';file-properties
"XNPEU";}}The weights are some positive integers which could be interpreted
as number of surfaces in the stack, Fig.6b). These branched surfaces inherit
from the train tracks some topological restrictions as depicted in Fig.7\ \
\ \FRAME{ftbpFU}{5.0618in}{2.1525in}{0pt}{\Qcb{Topological restrictions
which branched surfaces inherit from train tracks.}}{}{fig. 7.gif}{\special{
language "Scientific Word"; type "GRAPHIC"; maintain-aspect-ratio TRUE;
display "USEDEF"; valid_file "F"; width 5.0618in; height 2.1525in; depth
0pt; original-width 13.7185in; original-height 5.7917in; cropleft "0";
croptop "1"; cropright "1"; cropbottom "0"; filename 'C:/New Folder/fig.
7.gif';file-properties "XNPEU";}}\ \ \ \ \ \ \ \ \ \ \ \ \ \ \ \ \ \ \ \ \ \
\ \ \ \ \ \ \ \ \ \ \ \ \ \ \ \ \ \ \ \ \ \ \ \ \ \ \ \ \ \ \ \ \ \ \ \ \ \
\ \ \ \ \ \ \ \ \ \ \ \ \ \ \ \ \ \ \ \ \ \ \ \ \ \ \ \ \ \ \ \ \ \ \ \ \ \
\ \ \ \ \ \ \ \ \ \ \ \ \ \ 

For instance, in the case of train tracks the monogons are forbidden as well
as discs of contact, etc. (for details, please, contact Ref.[53]). Floyd and
Ortel [71] (see also Oertel [70 ], page 387) had proved the following

\bigskip

\textbf{Theorem 3.5.}\textit{There is a \textbf{finite} collection of
incompressible branched surfaces in some 3-manifold M}$^{{}}$\textit{such
that every 2-sided incompressible, }$\partial$\textit{-incompressible
surface in M}$^{{}}$\textit{\ is carried with positive weights by a branched
surface }$\frak{B}$\textit{\ of the collection.}

\bigskip

The surface $S$ is called 2\textit{-sided} in $M$ if there is a regular
neighborhood of $S$ homeomorphic to $S\times I$ where $I$=[0,1], [72]. If $%
S\neq S^{2}$, $RP^{2},$or a disc $D^{2}$ which can be pushed to $\partial M$%
, then $S$ is \textit{incompressible} if every loop (a simple closed curve)
on $S$ which bounds an (open) disc in $M\setminus S$ also bounds a disc in $S
$. If $S$ has boundary, then $S$ is also $\partial-incompressible$ if every
arc $\alpha$ in $S$ (with $\partial(\alpha)\in\partial S$) which is
homotopic to $\partial M$ is also homotopic in $S$ to $\partial S$ [19],
section 4.10.

Fig.6b) shows the local model for the fibered regular neighborhood $N$($%
\frak{B}$) of $\frak{B}$ , it also depicts the \textit{horizontal} boundary $%
\partial_{h}N(\frak{B})$ and the \textit{vertical }boundary $\partial_{v}N(%
\frak{B}).$

The branched surface $\frak{B}$ is incompressible if :

a) $\frak{B}$ has no discs of contact;

b) $\partial_{h}N(\frak{B})$ is incompressible and $\partial-$
incompressible in $M\setminus$\textit{\r{N} }where \textit{\r{N}} is

just an interior of $N$($\frak{B}$) and $\partial-$compressing disc for $%
\partial_{h}N(\frak{B})$ is assumed

to have boundary in $\partial M\cup\partial_{h}N(\frak{B});$

c)there are no monogons in $M\setminus$\textit{\r{N} .}

Being armed with such definitions, we would like to show how actually these
surfaces can be built.This task was actually accomplished in Refs.[73,74 ].
Alternative treatment can be found in recently published paper by Minsky
[47]. It is essential to contact these references for detailed study of this
very involved topic.The theory of incompressible surfaces in connection with
general theory of 3-manifolds can be found in the review by Jaco [75]. Here,
we restrict ourself \ only with very basic facts which are needed for
topological interpretation of the results which follow in section 4.

The mapping torus construction, Eq.(3.12), does produce 3-manifolds which
are the complements of the fig.8 and trefoil knots respectively. It is
clear, based on Eq.(3.19), that nothing much beyond this can be obtained. To
get more, one needs to generalize this construction. Following, Ref.[73], we
would like to consider more involved way of constructing torus bundles. For
example, let $S$ be some surface and let $X_{1},X_{2}$ be the topological
spaces associated with the mapping $\gamma_{i}:S\times\lbrack0,1]\rightarrow
X_{i}$ , i=1,2. For some surface homeomorphism $h$: $S\rightarrow S$ define X%
$_{1\text{ }}/_{h}X_{2}$ to be the quotient space obtained from $X_{1}\cup
X_{2}$ by identifying $\gamma_{1}(x,1)$ with $\gamma_{2}(h(x),0).$ In this
picture the mapping torus $T_{h}$, Eq.(3.12), is just a quotient space
obtained from $X_{1}$ by identification of $\gamma_{1}(x,1)$ with $%
\gamma_{1}(h(x),0).$ Now, however, we can extend this construction by
letting i to range from 1 to n.Thus, if $\gamma_{i}:$ $S\times\lbrack0,1]%
\rightarrow X_{i}$ and $h_{i}:$ $S\rightarrow S$ are homeomorphisms for
i=1,..., n, then the fiber bundle 
\begin{equation}
X=X_{1}/_{h_{1}}X_{2}/_{h_{2}}\cdot\cdot\cdot/_{h_{n-1}}X_{n}/_{h_{n}}  
\tag{3.36}
\end{equation}
is associated with some 3-manifold which, actually, can be obtained from the
figure 8 3-manifold by means of hyperbolic Dehn surgery [19,76]. Before
explaining its meaning, several simpler concepts need to be elucidated.
First, since according to section 2, all toral homeomorphisms can be
performed with help of the right $R$ and the left $L$ Dehn twists (for
orientation preserving homeomorphisms), the fiber bundle construction just
described is equivalent to some transformation $\varphi$ of the punctured
torus given by

\begin{equation}
\varphi=R^{a_{1}}L^{a_{2}}\cdot\cdot\cdot L^{a_{2n}}\text{ \ , \ a}%
_{i}\geq1,   \tag{3.37}
\end{equation}
in accord with Ref.[73]. Since each torus is completely determined by the
ratio $\tau$ of its sides, e.g. $\tau_{i}=a_{i}/b_{i}$ , it is clear, in
view of Eq.(3.36) that the transformation $\varphi$ should be such that $%
\varphi(a_{i}/b_{i})=a_{i+2n}/b_{i+2n}$ for all $i$ and fixed $n$ .This
prompts us to discuss the properties of the partition function $Z=tr\varphi$
which we postpone till section 4. In this section we would like to discuss a
bit more the topological issues. According to the results of section 2 the
transformation $\varphi$ is of the same kind as the transformation $W$ in
Eq.(2.50). This means that the cumulative result of matrix multiplications
in Eq.(3.37) is just some matrix \ $M$ of the type given by Eq.(2.31). In
section 2 we had already discussed special cases of equations for geodesics.
Here, we would like to investigate the most general case. Using Eq.(2.67) we
obtain the following roots which provide locations of the begining and the
end of \textbf{hyperbolic} geodesics on $S_{\infty}^{1}$%
\begin{equation}
x_{1,2}=\frac{1}{2c}(a-d\pm\sqrt{(a+d)^{2}-4}).   \tag{3.38}
\end{equation}
Since $a+d$ is just the trace of $M$ we have to impose the usual requirement
that $tr^{2}M\geq4$ for the transformation to be hyperbolic. In addition,
however, we have \ to take into account that $tr^{2}M$ is some nonnegative
integer. As we had argued in section 2, the roots $x_{1,2}$ \textbf{cannot}
be rational numbers. Hence, whatever they might be, they will belong to some
quadratic irrationalities. From the number theory [45 ] it is well known
that the continued fraction expansions of quadratic irrationalities are
periodic as it had been mentioned already after Eq.(2.58) and this fact also
explains the periodicity of $\varphi$ transformation.

\bigskip

\textbf{Remark 3.6}. Since, according to section 2, the exponents $%
a_{1},...,a_{2n}$\ in Eq.(3.37) are in one to one correspondence with the
coefficients of continued fraction expansion, this periodicity is in one to
one correspondence with the fiber bundles constructed in Eq.(3.36).

\bigskip

That this is indeed the case is shown in great detail in Ref. [73,74].
Moreover, since the transformation $\varphi$ is hyperbolic, it acts by
translations along the geodesics in $\mathcal{D}$ whose ends are determined
by the roots of Eq.(3.38). This has been illustrated already in Eq.s (2.60)
and (2.66). Imagine now that we move along one of such geodesics, say $\frak{%
G}$, that is \textbf{we move in the Teichm\"{u}ller space of the punctured
torus.} Such motion will be associated with the sequence of crossings of
triangles of the Farey tesselation. To understand why this is so the
folowing arguments are helpful. Each torus can be triangulated that is it
can be represented as at least two triangles. Obviously, one of this
triangles is sufficient for complete characterization of the torus. The
triangle is determined by the \textbf{slopes }of its sides. E.g. the
undistorted triangle is determined by the triple $\Delta(T)=<1/0,0/1,1/1>.$
It is sufficient to take a look at Fig.s 3 and 4 in order to recognize one
to one correspondence between this triple and the Farey numbers $\frak{F}_{1}
$ for the largest triangle in the Farey tesselation. Hence, motion in the
Teichm\"{u}ller space of the punctured torus is indeed associated with
sequence of triangles in the Farey tesselation of $\mathcal{D}$ since $M$($%
\Delta(T))=$ $\Delta(M(T))$ where $M(z)$ is defined by Eq.(2.37).
Topologically, we can visualize this chain of triangles as an infinite strip 
$\sum_{\varphi}$, e.g.see Fig.8, which consists of triangles crossed by $%
\frak{G}$.\FRAME{ftbpFU}{5.047in}{1.5843in}{0pt}{\Qcb{Bi infinite strip $%
\sum_{\protect\varphi}$ of edge-paths associated with hyperbolic
transformation $\protect\varphi$}}{}{fig.8.gif}{\special{ language
"Scientific Word"; type "GRAPHIC"; maintain-aspect-ratio TRUE; display
"USEDEF"; valid_file "F"; width 5.047in; height 1.5843in; depth 0pt;
original-width 9.6037in; original-height 2.9793in; cropleft "0"; croptop
"1"; cropright "1"; cropbottom "0"; filename 'C:/New
Folder/fig.8.gif';file-properties "XNPEU";}}The numbers a$_{i}>1$ indicate
the number of smaller triangles within each larger triangle.They are in one
to one correspondence with the exponents in Eq.(3.37). The periodicity of
the transformation $\varphi$ is reflected in the periodicity of the
triangulation pattern in the strip $\sum_{\varphi}.$To make all these
statements more physical, we need to discuss the meaning of hyperbolic Dehn
surgery (actually, Dehn filling!) now.

Let $M$ be a hyperbolic 3-manifold whose boundary $\partial M$ is a torus $T$%
. If we write $T$=$\mathbf{R}^{2}/\mathbf{Z}^{2}$and choose for basis
vectors \textbf{e}$_{1},\mathbf{e}_{2}$ of the lattice\textbf{\ Z}$^{2}$
then, for any coprime pair of integers (p,q) the element $p\mathbf{e}_{1}+q%
\mathbf{e}_{2}$ determines some simple curve on $T$. Consider another solid
torus $S^{1}\times D^{2}$. This torus we glue into $M$ in such a way that
the curve $p\mathbf{e}_{1}+q\mathbf{e}_{2}$ in $\partial M$ is being glued
to the meridian (that is to the curve of slope 1/0) of the solid torus.

\bigskip

\textbf{Remark 3.7}. It is being said that this new 3-manifold $M(p,q$) is
obtained from the old M by the operation of Dehn filling along the curve of
slope p/q.

\bigskip

\ Hatcher has proved the following remarkable theorem (using branched
surfaces) [77].

\bigskip

\textbf{Theorem 3.8}. \textit{A 3-manifold with a single torus boundary
component has only finitely many boundary slopes.}

\bigskip

\textbf{Remark 3.9 }a\textbf{)}.From the previous discussion it follows that
there should be a correspondence between the incompressible surfaces and the
boundary slopes.b) The boundary slopes are also in correspondence with 
\textbf{Q}$\cup\{\infty\}$ and, hence, with the Farey tesselation of $%
\mathcal{D}$.

\bigskip

This correspondence can be rephrased in physically familiar terms as we
would like to demonstrate now. To this purpose, let us recall that the
continued fraction expansion for $\alpha $ which is quadratic irrational can
be written as 
\begin{equation}
\alpha =[a_{0},...,a_{r},a_{r+1},...,a_{r+2n},a_{r+1},...a_{r+2n},...]. 
\tag{3.39}
\end{equation}
The quadratic irrational is \textbf{purely periodic} [78 ] if we have $%
a_{m+2n}=a_{m}$ for \textbf{all} $m$ (i.e. in this case, the terms $%
a_{0},...,a_{r}$ are absent ). Evidently, Fig.8 represents just this case.
Let us now associate the boundary slope p/q with continued fraction 
\begin{equation}
p/q=[a_{1},...,a_{2n}]  \tag{3.40}
\end{equation}
(e.g. see the discussion after Eq.(3.37) ) and consider \textbf{directed}
random walks from the point 1/0 to the point p/q on the diagram, Fig.8. The
random walk $\gamma $ is directed if it does not have backtracks, moreover,
the walk is \textbf{minimal} if two successive steps never belong to the
same triangle. A brief introduction to the properties of such walks can be
found in our earlier work, Ref.[79] , where such walk was used for
description of the discretized version of the Dirac propagator. Such
discretization is useful in some problems relevant to physics of
semiflexible polymers.

Following our earlier work, Ref.[79], let N=n be the total number of steps
in the directed(edge-paths) walk $\gamma $ and let N$_{+}$ be the number of
right turns while N$_{-}$ be the number of left turns in such a walk.
Associate a system of N Ising spins $\sigma _{i}$ with the walk so that $%
\sigma _{i}=+1(-1)$ will correspond to the turn to the left (right). With
such defined rule for spins, the ''magnetization'' \ \textbf{M} is given by 
\begin{equation}
\mathbf{M}=\sum\limits_{i=1}^{\text{N}}\sigma _{i}.  \tag{3.41}
\end{equation}
In the case if eigenvalues of the matrix $\varphi $ , Eq.(3.37), are
positive the boundary slope $m(n)$ is given by [74,80] 
\begin{equation}
m(n)=\frac{1}{4}\mathbf{M.}  \tag{3.42}
\end{equation}
If the matrix $\varphi $ has negative eigenvalues then, instead of
Eq.(3.42), one has $m(n)$=$(1/4)\mathbf{M+}1/2\mathbf{.}$ In the next
section thus introduced quantities acquire new statistical mechanical
meaning. \ Floyd and Hatcher [74] have proved the following theorem

\bigskip

\textbf{Theorem 3.10.}\textit{\ If }$\varphi$\textit{\ (Eq.3.37) is
hyperbolic, then a connected, orientable }

\textit{incompressible, }$\partial-$\textit{incompressible surface in M}$%
_{\varphi}$\textit{\ is exactly one of:}

\textit{a) the peripheral torus }$\partial M_{\varphi}$\textit{\ ,}

\textit{b) the fiber T}$^{2}-\{0\},$

\textit{c) a finite number (}$\geq2)$\textit{\ of \textbf{non-closed}
surfaces S}$_{\gamma}$\textit{\ indexed by minimal }

\textit{edge- paths }$\gamma.$

\bigskip

From here, it follows, in particular, that for  transformation given by the
matrix $a$, Eq.(2.21), which is relevant to 3-manifold associated with the
complement of fig. 8 knot, there are exactly 2 non-closed orientable
incompressible $\partial -$incompressible surfaces. According to Thurston
(Ref.[19], section 4.40) these are just $T^{2}\backslash 2$disks surfaces
which are oppositely oriented. The above Theorem 3.10 does not include the
nonorientable incompressible surfaces. This case was studied by Przytycki
[80,81] whose work extends the work by Floyd and Hatcher and is based on the
same edge-path methods. For the complement of fig.8 knot 3-manifold the
incompressible unoriented surface is just the Klein bottle-disk as it was
also shown by Thurston in the same section 4.40. We refer our readers to the
original papers for more details. In the meantime, we would like now to
provide some physical interpretation of the obtained results using some
recent results from statistical mechanics of number-theoretic spin chains.

\section{Thermodynamics of the Farey spin chains and statistical mechanics
of 2+1 gravity}

In our previous work, Ref.[16], we have used meanders and Peierls-type
arguments for description of phase transitions in liquid crystals and
gravity. Such type of approach, although provides some estimate of
transition parameters, is not well suited for more refined analysis since
even the notion of the order parameter, central to all theories of phase
transitions, is not so easy to implement within such an approach. In the
previous section we had introduced the logarithmic Mahler measure, Eq.
(3.24), which is ideally suitable for description of dynamical phase
transitions and in, addition, perfectly fits Yang and Lee theory of phase
transitions [7]. To make use of this measure in this section, we need to
provide an equivalent definition of this measure now (incidentally, this
alternative definition provides an extra link between the statistical
mechanics and the number theory [68]).To this purpose, instead of Eq.(3.23),
introduce 
\begin{equation}
\Delta _{n}(F)=\prod\limits_{i=1}^{d}(\alpha _{i}^{n}-1).  \tag{4.1}
\end{equation}
Using thus defined polynomial $\Delta _{n}$ it can be shown that, provided
no zero of $F(x)$, Eq.(3.22), is a root of unity, 
\begin{equation}
m(F)=\lim_{n\rightarrow \infty }\frac{1}{n}\ln \left| \Delta _{n}(F)\right| 
\text{ .}  \tag{4.2}
\end{equation}
In the case if polynomial $F(x)$ has zeros which are roots of unity, $m(F)=0.
$Being armed with such results, following Refs[82,83], let us introduce the
spin-like variable $\sigma _{i}=\{0,1\}$ , i=1-k , which is related to the
the Ising spin s$_{i}$=(-1)$^{\sigma _{i}}$ . Define now inductively the
matrix $D_{k}$ via the following set of rules: 
\begin{equation}
D_{0}=\left( 
\begin{array}{cc}
1 & 0 \\ 
0 & 1
\end{array}
\right)   \tag{4.3}
\end{equation}
and 
\begin{equation}
D_{k}=L^{1-\sigma _{k}}R^{\sigma _{k}}D_{k-1}(\sigma _{1},...,\sigma _{k-1}).
\tag{4.4}
\end{equation}
The energy $E_{k}$ can be defined now as 
\begin{equation}
E_{k}=\ln T_{k}  \tag{4.5}
\end{equation}
with $T_{k}=Tr(D_{k}).$ This allows us to introduce the partition function 
\begin{equation}
Z_{k}(\beta )=\sum\limits_{\{\sigma _{i}\}}\exp (-\beta E_{k}).  \tag{4.6}
\end{equation}
with control parameter $\beta $ playing a role of the inverse temperature.
The free energy $\frak{F}$($\beta )$ can be defined now in a usual way as 
\begin{equation}
\frak{F}(\beta )=\lim_{k\rightarrow \infty }F_{k}  \tag{4.7}
\end{equation}
where 
\begin{equation}
F_{k}=\frac{-1}{k\beta }\ln (Z_{k}(\beta )).  \tag{4.8}
\end{equation}
To make connection with Eq.(4.2) we notice that 
\begin{equation}
m(F)=\lim_{k\rightarrow \infty }\frac{1}{k}E_{k}\text{ .}  \tag{4.9}
\end{equation}
Since $E_{k}$ is random variable, more appropriate quantity is the average
Mahler logarithmic measure which can be interpreted as an average energy $%
U(\beta ),$ i.e. we have 
\begin{equation}
U(\beta )=\lim_{k\rightarrow \infty }U_{k}(\beta )\text{ with }U_{k}(\beta )=%
\frac{\partial }{\partial \beta }(\beta \mathit{F}_{k}(\beta )).  \tag{4.10}
\end{equation}
The average magnetization (per site) associated with the average boundary
slope, e.g. see Eqs.(3.41),(3.42), can be defined accordingly as 
\begin{equation}
M_{k}(\beta )=<\frac{1}{k}\sum\limits_{i=1}^{k}s_{i}>_{k}  \tag{4.11}
\end{equation}
where, as usual, \ for any observable $O$ the average $<...>_{k}$is defined
by 
\begin{equation}
<O>_{k}(\beta )=\frac{\sum\limits_{\{\sigma \}}O(\sigma )\exp \left( -\beta
E_{k}\right) }{\sum\limits_{\{\sigma \}}\exp \left( -\beta E_{k}\right) }%
\text{ .}  \tag{4.12}
\end{equation}
Detailed calculations performed in Refs.[7-10, 82, 83] allow us to avoid
repetitions. Hence, we only provide here the summary of results. In the
thermodynamic limit, k$\rightarrow \infty ,$ the partition function,
Eq.(4.6) acquires the following form: 
\begin{equation}
\hat{Z}(\beta )=\lim_{k\rightarrow \infty }Z_{k}(\beta )=\frac{\zeta (\beta
-1)}{\zeta (\beta )}  \tag{4.13}
\end{equation}
announced earlier in the Introduction, e.g. see Eq. (1.4).

\bigskip 

\textbf{Remark 4.1.} The partition function $\hat{Z}(\beta )$ can be brought
into form which coincides with that for lattice gases, e.g. see Eq.s(3.27)
and(3.28). To this purpose let us introduce the notation $\xi (\beta
)=\Gamma (\frac{\beta }{2})(\beta -1)\pi ^{-\frac{\beta }{2}}\zeta (\beta )$
where $\Gamma (x)$ is just the Euler's gamma function. Then, following
Riemann [2], we notice that 
\begin{equation*}
\xi (\beta )=\xi (0)\prod\limits_{\rho }(1-\frac{\beta }{\rho })
\end{equation*}
where $\rho $ ranges over the roots of the equation $\xi (\rho )=0$. Hence,
the complete statistical mechanics treatment of 2+1 gravity depends
crucially on our knowledge of distribution of zeros of Riemann zeta function
g($\rho )$ and, hence, on constructive solution of Riemann hypothesis.

\bigskip 

\textbf{Remark 4.2.} Taking into account that the Riemann's Zeta function
has the following integral presentation [1,2] 
\begin{equation*}
\zeta (\beta )=\frac{1}{\Gamma (\beta )}\int\limits_{0}^{\infty }dx\frac{%
x^{\beta -1}}{e^{x}-1}
\end{equation*}
one can interpret the obtained results in terms of the thermodynamic
properties of some fictitious Bose gas so that the phase transition in such
gas to some extent resembles Bose condensation. This topic is discussed
further in the Appendix A.1.

\bigskip 

Since solution of the Riemann hypothesis is not yet found (see, however,
Ref.[5]) we may be content for now by finding some meaningful estimates. For
instance, Eq. (4.13) allows one to obtain the critical ''temperature'' $%
\beta _{cr}=2$ [ 9\ ]. In the high temperature (pseudo-Anosov) phase, i.e.
for $0<\beta <\beta _{cr}$ , the averaged \ logarithmic Mahler measure $U$($%
\beta )$ is bounded by the following inequalities [9]: 
\begin{equation}
\frac{\ln 2-\beta \ln \frac{3}{2}}{2-\beta }\leq U(\beta )\leq \ln \frac{3}{2%
}.  \tag{4.14}
\end{equation}
Since for 1.7\TEXTsymbol{<}$\beta <2$ the l.h.s. of this inequality is
negative, more accurate estimate was obtained for 1$\leq \beta <2:$%
\begin{equation}
U(\beta )\geq \frac{1}{4}(\beta _{cr}-\beta ).  \tag{4.15}
\end{equation}
With such improved estimate Contucci and Knauff [9] had demonstrated that $%
U(\beta )\geq 0$ for 0$\leq \beta <2.$ Thus, the high temperature phase is
indeed of pseudo-Anosov type. For $\beta \geq \beta _{cr}$ , i.e. in the low
temperature (periodic or hexatic [20] ) frozen phase, $U$($\beta )=0$, which
is physically meanigful. In this phase also $\frak{F}$($\beta )=0.$ At the
same time, the average magnetization 
\begin{equation}
M(\beta )=\lim_{k\rightarrow \infty }M_{k}  \tag{4.16}
\end{equation}
is equal to one in the low temperature phase and is zero in the high
temperature phase.This also makes physical sense since, according to the
results of Przytycki, e.g. see Proposition 3.2. of Ref. [81], new $M(p,q)$
manifolds obtained by Dehn filling from 3-manifold $M$ which is complement
of the fig.8 knot are Seifert fibered only if p/q=$\pm 1/1,\pm 1/2,$or $\pm
1/3.$ In order to connect this result with physics it is important to
realize that the ''magnetization'' defined in Eq.(3.41) is made of spins
which are \textbf{periodically} arranged in the bi infinite strip $%
\sum\nolimits_{\varphi }.$ The thermodynamic averages are appropriate only \
if the period n$\rightarrow \infty .$ In the case of finite n some
n-dependent corrections are usually expected [20] for all observables. These
corrections naturally should vanish in the thermodynamic limit n$\rightarrow
\infty .$ Hence, identification of the average magnetization with the
average slope cannot be done automatically. This casts some doubts on the
appropriateness of the above thermodynamic formalism for description of
statistical mechanics of 2+1 gravity. Fortunately, there are other ways to
arrive at the same conclusions which we are now going to discuss in this and
the following section.

First, instead of considering the Farey triangulation of $\mathcal{D}$ we
can consider the associated with it dual tree as depicted in Fig.9. \FRAME{%
ftbpFU}{2.5071in}{2.5624in}{0pt}{\Qcb{Graph $\sum $ dual to the Farey
tesselation of $\mathcal{D}.$}}{}{fig.9.gif}{\special{language "Scientific
Word";type "GRAPHIC";maintain-aspect-ratio TRUE;display "USEDEF";valid_file
"F";width 2.5071in;height 2.5624in;depth 0pt;original-width
5.6351in;original-height 5.7605in;cropleft "0";croptop "1";cropright
"1";cropbottom "0";filename 'New Folder/fig.9.gif';file-properties
"XNPEU";}}\ Following Bowditch [35 ], we notice that, actually,
topologically such tree is in one to one correspondence with the Markov tree
depicted in Fig.2. This observation can be used for description of random
walks on such trees. This walk can be associated naturally with motions in
the Teichm\"{u}ller space of the punctured torus as has been observed
already by Penner [84]. Let $\sum $ denote the dual graph depicted in Fig.9
and let V($\tsum )$, E($\sum )$ and $\Omega $ denote respectively the sets
of vertices, edges and the regions complementary to $\sum $ . Each vertex
lies at the boundary of three complementary regions $X,Y,Z\in \Omega $ while
each edge meets four complementary regions $X,Y,Z,W$. The subset $\hat{\Omega%
}$ of nontrivial nonperipheral closed curves on T \ introduced in section 2
can now be identified with $\Omega $ because $X,Y,Z,W\in \Omega $ correspond
respectively to the equivalence classes of generators $a,b,ab$ and $ab^{-1}$
of the free group $G$ as discussed in section 2. Looking at the Fig.9 it is
easy to recognize that the regions of $\Omega $ are in one to one
correspondence with rationals \textbf{Q}$\cup \{\infty \}$ of the Farey
tesselation. Moreover, in the light of definitions just made, we can further
identify the Markov triples, $x$, $y$ and $z$ with the regions $X,Y,Z$ and $W
$ as follows : $x$=$\phi (X),etc.$ where $\phi (X)=tra,etc.$ Using Eq.s
(2.7),(2.8) one can easily obtain the following result 
\begin{equation}
z,w=h(x,y)=\frac{xy}{2}(1\pm \sqrt{1-4(\frac{1}{x^{2}}+\frac{1}{y^{2}})}). 
\tag{4.17}
\end{equation}
This results prompts us to rewrite Eq.s(2.7),(2.8) in the form 
\begin{equation}
\frac{x}{yz}+\frac{y}{xz}+\frac{z}{xy}=1  \tag{4.18}
\end{equation}
and 
\begin{equation}
\frac{z}{xy}+\frac{w}{xy}=1.  \tag{4.19}
\end{equation}
Written in such form, these equations acquire new geometrical meaning. Given
a directed edge $\vec{e}\in E(\sum )$ , define $\psi (\vec{e})=\frac{z}{xy}$
so that for each edge e we have, instead of Eq.(4.19), 
\begin{equation}
\psi (\vec{e})+\psi (-\vec{e})=1  \tag{4.20a}
\end{equation}
(edge relation), while instead of Eq.(4.18), we have 
\begin{equation}
\psi (\vec{e}_{1})+\psi (\vec{e}_{2})+\psi (\vec{e}_{3})=1.  \tag{4.20b}
\end{equation}
(vertex relation). It could be rather easily shown that $h(x,y)\geq
h(x)+h(y).$Given this fact, we observe that when 2$\leq \left| x\right| \leq
\infty $, $h(x)$ is real. Accordingly, we expect that $\left| x\right| $, $%
\left| y\right| ,\left| z\right| \geq 2.$Consider now some vertex $v^{\ast
}\in V(\sum )$ and consider the set $T_{n}(v^{\ast })$ \ which is a tree
spanned by all vertices which are at the distance at most $n$ from the
''seed''. Let $\Omega _{n}(v^{\ast })$ be the set of all complementary
regions meeting $T_{n}(v^{\ast })$ while $C_{n}(v^{\ast })$ be the
corresponding subset of edges. Then, it can be shown [85] that 
\begin{equation}
\sum\limits_{X\in \Omega _{n}(v^{\ast })}h(\phi (X))\leq \frac{1}{2}%
\sum\limits_{\vec{e}\in C_{n}(v^{\ast })}\psi (\vec{e})=\frac{1}{2} 
\tag{4.21}
\end{equation}
Actually, the inequality above can be replaced with equality and the above
equality is known as McShane identity[86]. Since, in spite of its
importance, we are not going to use it, we are not going to discuss its
significance. The above result is mentioned only because it is of interest
to inquire what happens with the convergence of some other functions $f(X).$
Let, for instance, $f$: $\Omega \rightarrow \lbrack 0,\infty )$ has a lower
Fibonacci bound (to be defined shortly below), then, according to Bowdich
[35], the series 
\begin{equation}
F(f)=\sum\limits_{X\in \Omega }[f(X)]^{-s}  \tag{4.22}
\end{equation}
converges for all s\TEXTsymbol{>}2. A lower Fibonacci bound for function $%
f(X)$on $\Omega _{n}\in \Omega $ exist if there is some constant k%
\TEXTsymbol{>}0 such that $f(X)\geq kF_{e}(X)$ for all but finitely many $%
X\in \Omega _{n}.$ The function $F_{e}(X)$ can be chosen as the length $%
\emph{L}$ of the reduced word W$_{r}$ [ 35\ ], e.g. see Eq.(2.5). Evidently, 
\emph{L}=$n$ in view of the results just presented. If \ this is so, the
question arises: how $n$ depends on $X$? Stated alternatively: is there way
to convert the summation over $X$ (that is over $\Omega _{n})$ into
summation over $n$? Fortunately, the last problem can be easily solved.
Since $X,Y,Z$ and $W$ are in one to one correspondence with the Farey
numbers all these numbers are coprime to each other [35]. Surely, that they
are also coprime to $n$.This means that the length $\emph{L}$ can take any
value n at precisely 2$\phi (n)$ regions of $\Omega _{n}$ where $\phi (n)$
has been defined by Eq.(1.5) as number of numbers less than n which are
prime to n. The factor of 2 is easy to understand if one puts $n$=1. Hence,
Eq.(4.22) can be rewritten as 
\begin{equation}
F(f)=2\sum\limits_{n=1}^{\infty }\phi (n)n^{-s}=2\frac{\zeta (s-1)}{\zeta (s)%
}  \tag{4.23}
\end{equation}
This result coincides with earlier obtained, Eq.(4.13), thus providing an
independent support to earlier developed thermodynamic formalism.

\section{Further developments}

\subsection{The circle packing and the Farey numbers}

Rademacher [48] had found alternative geomertic representation of the Farey
numbers which is worth discussing now since it has physical significance
which ultimately goes far beyond the leaky torus model. To facilitate our
reader's understanding, we would like to remind at this point few relevant
facts from the theory of M\"{o}bius transformations. To begin, using
Eq.(2.38) we notice that the point $\infty $ is the fixed point of the M\"{o}%
bius (in our case, modular) transformation as long as $c=0$. If this is the
case and, taking into account that $ad=1,$we obtain a transformation of the
type 
\begin{equation}
z^{\prime }=a^{2}z+ab  \tag{5.1}
\end{equation}
which will have another fixed point $z^{\ast }=ab/(1-a^{2})\neq \infty $ \
as long as $a\neq 1$. This cannot happen, however, as we had discussed
before Eq.(2.69). Therefore, the only possibility which is left to us is $a=1
$ and, hence, we are left with the parabolic transformation 
\begin{equation}
z^{\prime }=z+b.  \tag{5.2}
\end{equation}
This transformation fixes infinity and makes all integers equivalent (e.g.
discussion after Eq.(2.58)). It is directly associated with the presence of
a puncture (cusp) in the case of torus as can be easily proven [87]. In H$%
^{2}$ model realization of hyperbolic space consider the geodesic (or the
set of geodesics) which pass through point at infinity. Surely, these are
just semi infinite rays which are perpendicular to the real axis.
Accordingly, the associated horocycles are just  set of lines which are
parallel to the real axis. Consider, in particular, the horocycle which is
located at the vertical distance 1 from the real axis. The parabolic
transformation, Eq.(5.2), is \ not going to change the location of such
horocycle. But, in general, it is known, that the M\"{o}bius transformations
transform lines into lines and circles into circles. Hence, for the above
horocycle (actually a circle located at infinity) there must be a modular
transformation which transforms a circle at infinity to a circle located at
some point on the real axis. In particular, e.g.see Ref.[87], the
transformation $z$ $\longmapsto -1/z$ maps the horizontal horocycle at unit
height to the horocycle resting at the origin and having diameter 1. This
example can be easily generalized.To this purpose it is convenient to
rewrite the modular transformation, Eq.(2.37), \ as follows 
\begin{equation}
z^{\prime }-\frac{a}{c}=-\frac{1}{c^{2}(z+d/c)}.  \tag{5.3}
\end{equation}
Let now $Z^{\prime }=z^{\prime }-\dfrac{a}{c}$ and $Z=c^{2}(z+d/c)$ then, we
obtain back the transformation $Z^{\prime }=-1/Z.$ Let, now $z\ast =Z/c^{2}$
. By this transformation a circle of radius 1/2 becomes a circle of radius $%
1/2c^{2}$ touching the real axis at zero. Since c can be only an integer
number it is clear that all radii are less or equal to 1/2. Finally, by
using the transformation $z=z\ast -d/c$ we shift the point of tangency for
such circle to the location $z=-d/c$. Consider a very special case first:
c=1. Then, our circle is going to touch the real axis at some integer point
-d of real axis. It is clear now that for different d's we would have
different circles and that all these circles are going to touch each other.
Moreover, and this is not difficult to prove[98], for c different from 1 the
corresponding circles are all going to touch each other as it is depicted in
Fig.10.\FRAME{ftbpFU}{4.9692in}{2.1733in}{0pt}{\Qcb{Circle packing
associated with the Farey numbers.}}{}{fig.10.gif}{}Clearly, Fig.10 represents a
very coarse picture since the number of circles is countable infinity
between every neighboring pair (unit interval) of integers.Given this, one
may think about the distribution of sizes of such circles. Following
Sullivan [88], we say that two real numbers have the same $\rho -$size if
they belong to one of the intervals ($\rho ^{n+1},\rho ^{n}).$ Then, we can
group the circles into collections whose diameters have the same $\rho -$
size. The number of circles of a given size $s=(\rho ^{n+1},\rho ^{n})$
within a unit interval is the number of pairs $(p,q)$ with $p\leq q$ and $p$
relatively prime to $q$ (so that $s\sim 1/q^{2}$ ) which is again the Euler $%
\phi (q)$ totient function, Eq.(1.5). This fact can be easily understood if
we recognize, that the horocycles which are located at positions p/q within
unit interval are having sizes of order 1/q$^{2}.$

Following Sullivan [88] (and also Ref.[89]), let us consider definitions of
packing and covering Hausdorff measures. Adopted to our case, we have to
consider a (closed) subset $\Lambda $ of $R^{2}$ and to cover this subset by
discs of radii $r_{1},$ $r_{2}$, ..., all less than some $\varepsilon \geq 0.
$ Consider now the sum S$_{c}$ =$\sum\nolimits_{i}\psi (r_{i})$ where $\psi
(r_{i})=r_{i}^{\delta }$ with $\delta $ being some ''critical exponent''
(Hausdorff dimension) of the set $\Lambda .$ The \textit{covering} Hausdorff 
$\psi -$measure of $\Lambda $ is the limit (as $\varepsilon \rightarrow 0)$
of the \textit{infimum} of S$_{c}$ the exponent $\delta $ being a fractal
dimension of $\Lambda .$ Analogously, one can define the \textit{packing}
Hausdorff measure. In this case one should consider an open set $\Lambda
^{\prime }$ and to cover it by the set of disjoint (that is non touching)
circles. Then, one has to consider the \textit{supremum} for the analogous
sum S$_{p}$. In general the packing and the covering Hausdorff dimensions
are \textbf{not} the same. Such technicalities are needed for description of
limit sets of Fuchsian (M\"{o}bius in general) groups with cusps (for a
quick introduction to this subject, please consult our earlier work,
Ref.[25]). Depending upon the rank of the cusp one should use either S$_{c}$
or S$_{p}$ [90]. At this point it is just sufficient to take into account
that in our case, in view \ of the results just obtained, 
\begin{equation}
\sum\limits_{circles}(size)^{\frac{\delta }{2}}=\sum\limits_{n=1}^{\infty
}\phi (n)n^{-\delta }=\frac{\zeta (\delta -1)}{\zeta (\delta )}.  \tag{5.4}
\end{equation}
Hence, we have reobtained again the result, Eq.(4.13), which now acquires
completely new meaning. It should be apparent at this point, that one still
can do much better if one recalls all relevant facts about the
Patterson-Sullivan measure of the limit set \ $\Lambda $ which we had
discussed in our earlier work, Ref.[25], in connection with AdS/CFT
correspondence.

\subsection{\protect\bigskip The Eisenstein series and the S-matrix}

If $\rho(x,y)$ is the hyperbolic distance between points $x$ and $y$ $\in
H^{2}$ then, the Patterson -Sullivan measure can be constructed with help of
the Poincar$e^{\prime}$ series $g_{\delta}(x,y)$ defined as 
\begin{equation}
g_{\delta}(x,y)=\sum\limits_{\gamma\in\Gamma}\exp(-\frac{\delta}{2}%
\rho(x,\gamma y))   \tag{5.5}
\end{equation}
for some Fuchsian(or M\"{o}bius, in general) group $\Gamma.$ The factor $%
\delta$ is responsible for convergence/divergence of $g_{\delta}$ and its
threshold value is associated with the fractal dimension of the limit set $%
\Lambda$ (which is closed (sub)set of the boundary at infinity, in our case, 
$S_{\infty}^{1}).$ According to the theorem of Beardon and Maskit (Theorem
5.1. of Ref.[25]) the limit set $\Lambda$ of the discrete group $\Gamma$ is
made of parabolic limit points (i.e.those which are associated with cusps)
and conical limit points (i.e.those which are associated with the fixed
points of hyperbolic elements of $\Gamma$) and the conical limit points
always lie outside of the cusps.

This means that motion along the hyperbolic geodesics (e.g. see sections
2-4) is accompanied with countable infinity of events associated with such
geodesic entering and leaving the corresponding horocycle associated with
the cusp as it is schematically depicted in Fig.11. This fact was noticed in
the paper of Sullivan [89] who also had estimated the characteristic time of
this process (that is for the motion with unit speed along the geodesic
(world line) one can estimate how long such motion spends inside the cusp).
For additional illustration and more details, please, see Fig.5 of our
earlier work, Ref.[25], and the comments associated with it. \FRAME{ftbpFU}{%
5.0116in}{1.8965in}{0pt}{\Qcb{Entering and leaving the ''black hole'' (the
cusp) while moving along the hyperrbolic geodesic}}{}{fig.11.gif}{\special%
{language "Scientific Word";type "GRAPHIC";maintain-aspect-ratio
TRUE;display "USEDEF";valid_file "F";width 5.0116in;height 1.8965in;depth
0pt;original-width 13.7185in;original-height 5.1456in;cropleft "0";croptop
"1";cropright "1";cropbottom "0";filename 'C:/New
Folder/fig.11.gif';file-properties "XNPEU";}}Let our geodesic begins at the
point z* and ends at the point w* \ of real axis. These points are chosen in
such a way that it passes through the point z which is the top of horocycle
located at the point -d/c of real axis and through the point $w$ of the
horocycle located at infinity. Hence, the point $w=i+x$, the point $z=iy-d/c$
(and $y=1/$c$^{2}).$ It is always possible to find a transformation $g(z)$
such that $g(z\ast )=0$, $g(w\ast )=\infty $, $g(z)=iy$ and $g(w)=i$ [91].
In view of this, we obtain, 
\begin{equation}
\rho (z,w)=\ln (\frac{1}{y}).  \tag{5.6}
\end{equation}
Using this result in Eq.(5.5) we obtain, 
\begin{equation}
E(y,\delta )=\sum\limits_{\gamma \in \Gamma }(\gamma y)^{\frac{\delta }{2}}.
\tag{5.7}
\end{equation}
This is just the Eisenstein series. Evidently, the subset $\Gamma $ must
correspond to subset of closed \textbf{nonperipheral} curves (that is, it
does not contain an accidental parabolic elements) on the torus which belong
to $\hat{\Omega}$ (defined after Eq.(2.5)). Fortunately, the properties of
the Eisenstein series, Eq.(5.7), are well known. This fact allows us not
only to provide different interpretation to our main result, Eq.(4.13), but
allows, in principle, to generalize the obtained results to the Riemann
surfaces of higher genus. To understand why this is so, we would like to
reproduce some results from our previous work, Ref[25], at this time.

In the upper half space model realization of d+1 dimensional hyperbolic
space H$^{d+1}$the hyperbolic Laplacian $\Delta _{h}$ acts on some function $%
f(\mathbf{x},z)$ according to the following prescription 
\begin{equation}
\Delta _{h}f(\mathbf{x},z)=z^{2}\left[ \Delta f-(d-1)\frac{1}{z}\frac{%
\partial f}{\partial z}\right] \text{ , }z>0\text{, \textbf{x}}\in \mathbf{R}%
^{d}.  \tag{5.8}
\end{equation}
In 2 dimensions the second term vanishes. We would like to keep it,
nevertheless, since the obtained results can be immediately generalized (see
below). For any $d\geq 1$ the eigenvalue equation for the hyperbolic
Laplacian reads: 
\begin{equation}
\Delta _{h}z^{\frac{\delta }{2}}=\frac{\delta }{2}(\frac{\delta }{2}-d)z^{%
\frac{\delta }{2}}.  \tag{5.9}
\end{equation}
In the case of two dimensions, of course, we have to replace $z$ by $y$. If 
\begin{equation}
\Delta _{h}f(x)\equiv F(x)  \tag{5.10}
\end{equation}
and x=\{\textbf{x},z\} then, for any $\gamma \in \Gamma $ where $\Gamma $ is
the group of isometries which leave H$^{d+1}$ invariant, we obtain 
\begin{equation}
\Delta _{h}f(\gamma x)\equiv F(\gamma x).  \tag{5.11}
\end{equation}
This means that not only $z^{\frac{\delta }{2}}$ is an eigenvalue of $\Delta
_{h\text{ }}$but $\left( \gamma z\right) ^{\frac{\delta }{2}}$ as well.Of
course, the linear combination is also an eigenvalue of $\Delta _{h}$. Thus,
not only for d=1 \ but \textbf{for any} d$\geq 1$ we obtain, 
\begin{equation}
\Delta _{h}E(z,\delta )=\frac{\delta }{2}(\frac{\delta }{2}-d)E(z,\delta ). 
\tag{5.12}
\end{equation}
Surely, for $d=1$ we relabel $z$ as $y$. In this case, following [92] we
obtain, 
\begin{equation}
E(y,\delta )=y^{\frac{\delta }{2}}+\varphi _{\Gamma }(\delta )y^{1-\frac{%
\delta }{2}}+O(e^{-2\pi y}),  \tag{5.13}
\end{equation}
which is sufficient for large y's. Here the scattering S-matrix $\varphi
_{\Gamma }(\delta )$ [12] is given by 
\begin{equation}
\varphi _{\Gamma }(\delta )=\sqrt{\pi }\frac{\Gamma (\frac{\delta }{2}-\frac{%
1}{2})\zeta (\delta -1)}{\Gamma (\frac{\delta }{2})\zeta (\delta )} 
\tag{5.14}
\end{equation}
This result should be compared with Eq.s(4.13) and (5.4). In order to
generalize these results we would like to mention several useful properties
of the S-matrix $\varphi _{\Gamma }(\delta )$. Fist, it can be shown [92],
that 
\begin{equation}
E(y,\delta )=\varphi _{\Gamma }(\delta )E(y,1-\delta ).  \tag{5.15}
\end{equation}
Using this result, we introduce new variable $\delta -\frac{1}{2}=\xi .$
Eq.(5.15), when written in terms of this new variable, acquires the
following more symmetric form 
\begin{equation}
E(y,\xi +\frac{1}{2})=\varphi _{\Gamma }(\xi +\frac{1}{2})E(y,\xi -\frac{1}{2%
}).  \tag{5.16}
\end{equation}
Let now $\xi \rightarrow -\xi $ in Eq.(5.16). By combining thus obtained
equation with Eq.(5.16) we obtain very important relation (unitarity
condition) [93]. 
\begin{equation}
\varphi _{\Gamma }(\xi +\frac{1}{2})\varphi _{\Gamma }(\xi -\frac{1}{2})=1. 
\tag{5.17}
\end{equation}
Being armed with this results we can now proceed with generalizations.

\subsection{Hyperbolic Dehn surgery once again}

In this work we only provide an outline of the relevant results leaving more
detailed discussion for future publications. To begin, let us notice that so
far we were able to obtain \textbf{all} results of this paper only because
the Teichm\"{u}ller space of the punctured torus happen to coincide with the
hyperbolic upper plane Poincar$e^{\prime }$ model H$^{2}$so that the motion
in the Teichm\"{u}ller space \textbf{coincides} with the motion in the
hyperbolic space H$^{2}$. \ This is not the case for the Riemann surfaces of
higher genus [41] and, hence, at first sight, the results of this paper are
not extendable to surfaces of higher genus. Very fortunately, this is not
the case as we would like now to argue. Let us recall, that we have started
our discussion with the figure eight and the trefoil knots in section 3.
Then, we had considered the corresponding 3-manifolds created by the mapping
torus construction and, after this, we had considered the incompressible
surfaces (in section 3.3). Three-manifolds associated with these surfaces
had been obtained with help of the Dehn surgery (Dehn filling) from the
''parent'' 3-manifold (e.g. see Remark 3.7) which is just that for the
figure eight complement in $S^{3}.$ Neumann and Zagier in the benchmark
paper [94] had developed a sort of perturbative calculations which allow to
estimate volumes of 3-manifolds obtained from the parent 3-manifold by the
operation of Dehn filling. In particular, for the decendants of the figure
eight hyperbolic manifolds they obtained the following estimate for the
volume of M(p,q) 3- manifold 
\begin{equation}
Vol(M(p,q))=Vol(M_{8})-\frac{2\sqrt{3}\pi ^{2}}{p^{2}+12q^{2}}+\frac{4\sqrt{3%
}(p^{4}-72p^{2}q^{2}+144q^{4})\pi ^{4}}{3(p^{2}+12q^{2})^{4}}+...  \tag{5.18}
\end{equation}
where $Vol(M_{8})$ is known to be 2.0298832... The obtained result is the
simplest in the chain of results obtained in this reference. In general, $%
Vol(M(p,q))<Vol(M)$ where $M$ is parental 3-manifold. This constitutes the
essence of Thurston's Dehn surgery theorem [19,76].

In our previous work, Ref.[25], we had considered 3 manifolds with cusps, in
particular, the 3-manifold for the figure eight knot contains just one 
\textbf{Z}$\oplus \mathbf{Z}$ cusp. It can be shown, e.g. see the Appendix
A.3, that \textbf{all} hyperbolic 3-manifolds associated with knots and
links are \textbf{Z}$\oplus \mathbf{Z}$ cusped: one cusp for \textbf{each}
embedded circle $S^{1}.$ Hence, such manifolds are necessarily \textbf{%
noncompact} but, nevertheless, of\textbf{\ finite} volume. Surely, there are
3-manifolds without cusps too and these are related to those with cusps.
According to Thurston (e.g. see section 5.33 of [19] ), all 3-manifolds $%
M(p,q)$ obtained by Dehn filling are \textbf{without} cusps. And, moreover, 
\begin{equation}
\lim_{(p,q)\rightarrow \infty }Vol(M(p,q))=Vol(M).  \tag{5.19}
\end{equation}
Obtained result is naturally extendable to the case of $k$ cusped manifolds.
By introducing notations 
\begin{equation*}
M_{k}=M_{(p_{1,}q_{1,}...\text{ ,\ }p_{k,}q_{k})}
\end{equation*}
the following general result was obtained by Neumann and Zagier [94] 
\begin{equation}
Vol(M_{k})=Vol(M)-\frac{\pi }{2}\sum\limits_{i=1}^{k}L_{i}+O(L_{i}^{2}) 
\tag{5.20}
\end{equation}
where $L_{i}$ is the length of short geodesic $\gamma _{i}$ on $M_{k}$ which
is just the length of the core curve of the solid torus added at i-th cusp.
The result of major importance for us is the observation by Neumann and
Zagier that the \textbf{difference of volumes depends to a high order only
on the geometry of} \textbf{cusps} and \textbf{not} on the rest of $M$ [94].
Given this observation, the following question can be asked immediately : is
it possible to reobtain the results Eq.s(5.13)-(5.15) for cusped 3-
manifolds ? The answer is YES! if these manifolds belong to the arithmetic
hyperbolic 3-manifolds. We provide a condensed summary of results related to
such manifolds in the Appendices A.2 and A.3 while in the main text we
discuss the scattering theory for such manifolds.

\subsection{Scattering theory for arithmetic 3-manifolds}

Let is begin with observation that Eq.(5.12) is valid for any d$\geq 1.$
Accordingly, the asymptotic expansion, Eq.(5.13), also survives and acquires
the following form (for H$^{3})$ 
\begin{equation}
E(y,\delta )=y^{\frac{\delta }{2}}+\Phi _{\Gamma }(\delta )y^{2-\frac{d}{2}%
}+O(e^{-cy})  \tag{5.21}
\end{equation}
with c being some constant.The question immediately arises: will the
S-matrix $\Phi _{\Gamma }(\delta )$ have the same form as in Eq.(5.14)? In
general, the answer is ''NO'' but for the arithmetic 3-manifolds the answer
is ''YES'' as had been demonstrated by Sarnak[26]. It is instructive to
restore some of his calculations now. In order to do so, we have to think
about the ways the group PSL(2,Z) can be extended in order to be used for
description of 3-manifolds associated with the punctured torus bundle.

To this purpose we notice that the group PSL(2,Z) is just a subgroup of
PSL(2,R) which is group of isometries of H$^{2}.$The group of isometries of H%
$^{3}$ is PSL(2,C) as is well known [25]. Hence, we have to figure out what
kind of subgroup of PSL(2,C) is analogous to PSL(2,Z). \ This problem was
actually solved for the torus bundles in Ref. [95] and for surfaces of
higher genus in very important paper by Margulis [30] to be discussed
further in Appendix A.3. For the torus bundles, naturally, one should try to
find solutions of Eq.(2.7) in the complex domain. In Ref.[95] it was shown
that the complex counterpart of the triple (3,3,3) is 
\begin{equation}
(\frac{3+\sqrt{-3}}{2},\frac{3+\sqrt{-3}}{2},\frac{3-\sqrt{-3}}{2}) 
\tag{5.22}
\end{equation}
and the rest of triples also can be obtained . The numbers above are
integers which belong to the ring of integers $\mathcal{O}_{3}$ in the
quadratic number field \textbf{Q}($\sqrt{-3}$) ,e.g. read Appendix A.2.
Hence, for the punctured torus bundle the analog of PSL(2,Z) is PSL(2,$%
\mathcal{O}_{3})$ and the punctured torus fiber bundle is 3-orbifold H$%
^{3}/PSL(2,\mathcal{O}_{3})$ as was proven rigorously in Ref.[95]. The
question remains: can this result be generalized to the noncompact (that is
having punctures (or cusps)) 3-manifolds originating from automorphisms of
Riemann surfaces of genus higher than one? The answer is ''YES'', provided
that discrete group of isometries is Bianchi group PSL(2,$\mathcal{O}_{d}),$
where 
\begin{equation}
-d=1,2,3,5,6,7,10,11,14,15,19,23,31,35,39,47,55,71,95,119.  \tag{5.23}
\end{equation}
This remarkable result can be found in Ref.[96]. What is even more
remarkable, that it is just a refinment of the very comprehensive work by
Bianchi [28] which was completed already in 1892 ! In Ref.[97], Proposition
1, Reid had proved the following theorem

\bigskip

\textbf{Theorem 5.1.}\textit{Every non-compact arithmetic Kleinian group is
conjugate in PSL(2,C) to a group commensurable with some Bianchi group
PSL(2, }$\mathcal{O}_{d}).$

\bigskip

\textbf{Remark 5.2.}a\textbf{)}The definition of arithmeticity is rather
involved and is discussed in the Appendices A.2 and A.3. The above theorem
is complementary to the theorem by Margulis ( e.g. Theorem A.3.11 of
Appendix A.3.) which is much more comprehensive.b) The noncompactness
implies existence of the parabolic peripheral subgroups (e.g. see section
2).c)The notion of commensurability between groups can be found, for
example, in Ref.[98], and is associated with set theoretical definition of
overlap (i.e. $\cap )$ between sets.

\bigskip

\textbf{Corollary 5.3.} \textit{Every cusped arithmetic hyperbolic
3-orbifold is commensurable with Bianchi orbifold H}$^{3}/PSL(2,\mathcal{O}%
_{d})$\textit{\ where }$\mathcal{O}_{d}$\textit{\ is ring of integers in the
quadratic number field \textbf{Q}(}$\sqrt{-d}$\textit{).}

\bigskip

\textbf{Remark 5.4}.\textit{\ }a)There is one-to-one correspondence between
the rings S$^{1}$\ in some(arithmetic) link and the cusps (e.g.read
AppendixA.3. for more details). b) The non-compactness is caused by cusps.c)
Cusped 3-manifolds do have finite volume(e.g. read section 5.3 again). d)
The arguments of section 5.1.strongly suggest that 3-manifold associated
with the figure eight knot is arithmetic. This is rigorously proven by Reid
in [97] who also proved that the figure eight is the only knot which is
arithmetic.e) Corollary 5.3 and Eq.(5.23) suggest that there is a countable
infinity of arithmetic links.

\bigskip

Being armed with these results we can now discuss the results of Sarnak
[26]. To this purpose, we have to rewrite Eq.(5.7) in a more convenient (for
the present purposes) form.We have ($z=x+iy$): 
\begin{equation}
E(y,\delta )=y^{\frac{\delta }{2}}+\sum\limits_{c=1}^{\infty
}\sum\limits_{(c,d)=1}\sum\limits_{m=-\infty }^{\infty }\frac{1}{\left[
\left( c(x+m)+d_{0}\right) ^{2}+c^{2}y^{2}\right] ^{\frac{\delta }{2}}} 
\tag{5.25}
\end{equation}
e.g. see Lax and Phillips [12], pages.171-172. Here in the second sum
summation is over d provided that c and d are relative primes, i.e. $(c,d)=1$%
, $d$=$d_{0}$+$mc$ and 0$\leq d_{0}$ \TEXTsymbol{<}$c$. The summation
procedure is explained in great detail in the same reference. In the
Sarnak's case one needs only to replace $x$ by $z$ and $y$ by $t$ (t 
\TEXTsymbol{>}0) (in the H$^{3}$ model realization of hyperbolic space).The
details of calculations can be found in Ref.[26] so that the final result,
indeed, has the form given by Eq.(5.21) where for the quadratic number field 
$\mathbf{Q}(\sqrt{-d})$ of class number one(e.g. read Appendix A.2. for
explanation of terminology) $\Phi _{\Gamma }(\delta )$ is given by 
\begin{equation}
\Phi _{\Gamma }(\delta )=\frac{\pi }{V(F_{L})(\frac{\delta }{2}-1)}\frac{%
\zeta _{d}(\frac{\delta }{2}-1)}{\zeta _{d}(\frac{\delta }{2})}  \tag{5.26}
\end{equation}
with $V(F_{L})$ being defined as 
\begin{equation}
V(F_{L})=\left\{ 
\begin{array}{c}
\sqrt{D}/2\text{ \ if }D=3\text{(}mod\text{4)} \\ 
\sqrt{D}\text{ if D}\neq 3(\func{mod}4)
\end{array}
\right.   \tag{5.27}
\end{equation}
provided that 
\begin{equation*}
d=\left\{ 
\begin{array}{c}
-D\text{ \ if D=3(mod4)} \\ 
-4D\text{ \ if D}\neq 3(\func{mod}4)
\end{array}
\right. .
\end{equation*}
Since, according to Appendix A.2, Eq.(A.2.12), the Dedekind zeta function $%
\zeta _{d}(s)$ has the same pole as the ordinary zeta function, the result,
Eq.(5.21), can be used instead of earlier obtained Eq.(5.14). This time,
however, one can get much more. In particular, one can \ get the volume $%
V(F_{D})$ of the associated cusped hyperbolic 3-manifold. The expression for
the volume is known to be [19,26,98] : 
\begin{equation}
V(F_{D})=\frac{\left| d\right| ^{\frac{3}{2}}\text{ }\zeta _{d}(2)}{4\pi ^{2}%
}.  \tag{5.28}
\end{equation}
This result can be obtained directly from the Eisenstein series,Eq.(5.25).
It comes as the residue at the pole $\frac{\delta }{2}=2$ of the Dedekind
zeta function in the numerator Eq.(5.26) [26]. That is 
\begin{equation}
\func{Re}\mathit{s}[E(y,\delta ),\frac{\delta }{2}=2]=\frac{V(F_{L})}{%
V(F_{D})}.  \tag{5.29}
\end{equation}
Since both the residue and $V(F_{L})$ are known, the volume, Eq.(5.28), is
obtained using Eq.(5.29).

Consider now generalization of the obtained results to the case of multiple
cusps (that is to the arithmetic links (e.g. see Appendix A.3). Following
Efrat and Sarnak [27 ],and also [98], consider a complete set \{ $\kappa
_{1}=\infty ,\kappa _{2},...,\kappa _{n}$ \} of the nonequivalent cusps and
let $\Gamma _{i}$ be a subgroup of $\Gamma \subset PSL(2,C)$ for which $%
\kappa _{i}$ is the fixed point that is $\Gamma _{i}$ \ is stabilizer in $%
\Gamma $ of the i-th cusp. It is always possible to select transformation $%
\rho _{i}(\kappa _{i})=\infty $ where $\rho _{i}\in \Gamma .$Then, 
\begin{equation}
\rho _{i}\Gamma _{i}\rho _{i}^{-1}=\left( 
\begin{array}{cc}
1 & \omega  \\ 
0 & 1
\end{array}
\right)   \tag{5.30}
\end{equation}
which coincides with the matrix $U_{1}$ defined in Eq.(A.3.1) with $\omega $
being a unit of quadratic number field. For each cusp one can define its own
Eisenstein series analogous to Eq.(5.7). Also, by analogy with Eq.(5.21) we
write now 
\begin{equation}
E_{i}(y,\delta )=\delta _{i,j}y^{\frac{\delta }{2}}+\Phi _{i,j}(\delta )y^{2-%
\frac{\delta }{2}}\text{ , i,j=1,...,n .}  \tag{5.31}
\end{equation}
It can be shown, that the $S-$matrix $\Phi _{i,j}$ obeys the matrix equation 
\begin{equation}
\mathbf{E}(y,\delta )=\mathbf{\Phi }(\delta \mathbf{)E}(y,2-\delta ) 
\tag{5.32}
\end{equation}
analogous to Eq.(5.15) discussed earlier. Finally, the analog of the
Eq.(5.14) is the determinant of the matrix $\mathbf{\Phi (}\delta \mathbf{)}$
given by 
\begin{equation}
\det \mathbf{\Phi }(\delta \mathbf{)=}const\frac{\xi _{H}(\frac{\delta }{2}%
-1)}{\xi _{H}(\frac{\delta }{2})}  \tag{5.33}
\end{equation}
to be compared with Eq.(5.26). $\xi _{H}(s)$ for the quadratic number field
of class number one is given by 
\begin{equation}
\xi _{H}(s)=\left( \sqrt{d}/2\pi \right) ^{s}\Gamma (s)\zeta _{H}(s) 
\tag{5.34}
\end{equation}
with $\zeta _{H}(s)$ being defined by 
\begin{equation}
\zeta _{H}(s)=\prod\limits_{\chi }L(s,\chi ).  \tag{5.35}
\end{equation}
It should be noted, that the above product over characters (which are
specific for each quadratic field) always starts with $\chi =1$ . This
produces $L(s,1)=\zeta (s)$ where, again, $\zeta (s)$ is usual Riemann zeta
function. Hence, again, the singularities of the determinant are just those
for the ratio of zeta functions as in Eq.(5.4). Hence, the exact partition
function for 2+1 gravity is given by the determinant det$\mathbf{\Phi }$,
Eq.(5.33), of the scattering S-matrix $\mathbf{\Phi }.$

\bigskip

\textbf{Acknowledgments}\ \ \ \ \ The author would like to thank Jeff Weeks
(Geometry Center U.Minnesota) for encouragement and many useful references,
Josef Przytycki (G.Washington University) for sending his papers and the
unpublished manuscript of T.Jorgensen, Bill Floyd (Virginia Tech) for many
clarifying remarks related to the incompressible surfaces, Alan\ \ Reid (U.
of Texas at Austin) for sending chapters of his unpublished yet book on
arithmetic hyperbolic manifolds, Scot Adams (U. of Minnesota) for\ sending
his papers related to Lorentz manifolds and arithmeticity and sharing the
information about his planned book on this subject.\ \ \ \ \ \ \ \ \ \ \ \ \
\ \ \ \ \ \ \ \ \ \ \ \ \ \ \textbf{\ \ \ }

\bigskip\pagebreak

\textbf{\ }

\bigskip

\textbf{Appendix A.1}

\ \textbf{\ }

\textbf{Phase transitions in 2+1 gravity:\ \ analogy with\ Bose\ condensation%
}

\bigskip

Although in section 4 the limiting value of the partition function,
Eq.(4.13), is obtained in a systematic way [9,10], the extraction of useful
information from this result is plagued by some serious problems as was
noticed already in earlier work of Cvitanovic [15 ]. The result, Eq.(4.13),
is obtained as limiting value of the Farey spin chain partition function
whose size tends to infinity. Since in Section 5 we generalize the punctured
torus results to surfaces of higher genus, the analogy with spin chain
cannot be straightforwardly extended. Hence, we need to develop methods of
extraction of useful information without being \ dependent on spin chain
analogy.

The main difficulty with Eq.(4.13) lies in the fact that it contains
singularities.This can be easily demonstrated if we notice that for large
values of $n$ the Euler totient function $\phi (n)\sim n$ [45]. This
observation allows us to write 
\begin{equation}
\hat{Z}(\beta )\leq \zeta (\beta -1).  \tag{A.1.1}
\end{equation}
Since $\zeta (\beta -1)$ is singular at $\beta _{cr}=2$ one is faced with
the problem of removing this singularity in physically acceptable way. There
are many ways of doing so but it is not our purpose here to provide a
complete list of possibilities. Rather, we would like to notice the
following. First, by direct numerical calculation one observes that for
large $\beta ^{\prime }s$ the partition function $\hat{Z}(\beta )$
approaches one. So that, indeed, for low ''temperatures'' the free energy is
zero in accord with earlier results [9,10]. The authors of these papers
claim that $\hat{Z}(\beta )$ is zero as soon as $\beta >2.$ This can be
understood only if one considers finite number terms in the corresponding
zeta functions and takes the thermodynamic limit using Eq.(4.8). To be
consistent, we need to do the same for $\beta <2.$ This is not a simple task
however since, according to Ref.[10] (e.g. see p.428), ''For high
temperatures ($\beta <2)$ the analytic continuation of $\hat{Z}(\beta )$ 
\textbf{cannot} be directly interpreted as the partition function of the
infinite chain''. This statement is rather pessimistic, especially, if one
is thinking about extension of obtained results to surfaces of higher genus.
Therefore, we offer here a somewhat different interpretation of the obtained
results.

To this purpose, let us introduce new function $g_{\beta }(\alpha )$ defined
by the following equation: 
\begin{equation}
g_{\beta }(\alpha )=\sum\limits_{n=1}^{\infty }\frac{\alpha ^{n}}{n^{\beta }}%
=\frac{1}{\Gamma (\beta )}\int\limits_{0}^{\infty }dx\frac{x^{\beta -1}}{%
\alpha ^{-1}e^{x}-1}.  \tag{A.1.2}
\end{equation}
Such type of functions are well known from the theory of Bose-Einstein (B-E)
condensation [99].The analogy, unfortunately, is not complete since in the
case of B-E condensation the exponent $\beta -1$ is never zero or less than
zero. Evidently, for $\alpha =1$ we obtain back the Riemann's zeta function.
The parameter $\alpha $ is related to the chemical potential $\mu $ in a
usual way. For $\mu =0$ we have $\alpha =1$. This value of $\alpha $ is
associated with singularity which has physical interpretation. In properly
chosen system of units equation which connects the chemical potential with
the particle density $\rho =\frac{N}{V}$ in B-E case reads [99]: 
\begin{equation}
\rho =g_{\frac{3}{2}}(\alpha )+\frac{1}{V}\frac{\alpha }{1-\alpha } 
\tag{A.1.3}
\end{equation}
where $V$ is volume of the system and $N$ is the total number of
particles.The second term in the r.h.s. represents the density of Bose
condensate associated with zero translational mode. For finite volume this
density is infinite for $\alpha =1.$ Therefore, as usual, it is assumed that
both the volume and the density tend to infinity in such a way that the
ratio is finite number. In our case, we do not have the luxury of having
volume term but we can extract the singularity in a manner similar to the
B-E gas.Taking into account Eq.s (A.1.1) - (A.1.3) we can subdivide the
domain of integration into two parts : a) 0$\leq x\leq 1$ and b)1$\leq x\leq
\infty .$ In the first case we obtain, 
\begin{equation}
I_{1}=\int\limits_{0}^{1}dxx^{\beta -3}=\frac{1}{\beta -2},  \tag{A.1.4}
\end{equation}
which is expected singularity of the zeta function at already known critical
value $\beta _{cr}=2.$ As for the second part we obtain (close to
criticality) 
\begin{equation}
I_{2}=\int\limits_{1}^{\infty }dx\frac{x^{\beta -2}}{e^{x}-1}\simeq
\int\limits_{1}^{\infty }dx\frac{1}{e^{x}-1}+\left( \beta -2\right)
\int\limits_{1}^{\infty }dx\frac{\ln x}{e^{x}-1}+...  \tag{A.1.5}
\end{equation}
Combining Eqs.(A.1.4) and (A.1.5) we obtain as well, 
\begin{equation}
I=\frac{1}{\varepsilon }(1+c_{1}\varepsilon )(1+c_{2}(\beta )\varepsilon
^{2}+...),  \tag{A.1.6}
\end{equation}
where $\varepsilon =\beta -2$ and c$_{1}$,c$_{2}(\beta )$ are known
constants. In particular, $0.373<c_{2}<0.5438$ for 1$\leq \beta \leq 2.$%
Taking into account that $\Gamma (\beta )$ is nonsingular in the range of $%
\beta ^{\prime }s$ which is physically interesting, we can subtract the
singular part along with well behaving regular part in order to get 
\begin{equation}
-\beta \frak{F(\beta )=}ln(1+c_{2}(\beta )\varepsilon ^{2}+...)\geq
c_{2}(\beta )\varepsilon ^{2}.  \tag{A.1.7}
\end{equation}
This result coincides almost exactly with the estimate, Eq.(25), obtained in
Ref.[9] (the constant c$_{2}$ is $\approx 0.25$ in Ref.[9]) for the free
energy in the high temperature ''disordered'' (or pseudo-Anosov) phase.
Hence, the rest of arguments of Ref.[9] now can be used. In particular, we
have 
\begin{equation}
-\beta \frak{F(\beta )=}\int\limits_{\beta }^{2}U(x)dx\leq U(\beta )(2-\beta
).  \tag{A.1.8}
\end{equation}
Using this result, we obtain, 
\begin{equation}
U(\beta )\geq --\beta \frak{F(\beta )}/(2-\beta )\geq c_{2}(\beta )(2-\beta
),\text{ \ \ 1}\leq \beta \leq 2,  \tag{A.1.9}
\end{equation}
in complete accord with Eq.(4.15) of the main text.

\bigskip 

\textbf{Remark A 1}.\textbf{1}. a) The analogy with Bose gas condensation
can be strengthened, perhaps, even more if one notices that the criticality
condition $\beta _{cr}=2$ coincides with the simple estimate made in famous
Kosterlitz and Thouless paper [23].In this paper the Coulomb gas model is
used for description of phase transitions in two dimensional liquid
crystals, liquid helium, etc. This, in part, explains the success of such
type of models prevailing so far in physics literature [20].b) In the case
of liquid helium Feynman [100] had obtained equation very similar to
Eq.(A.1.3) and Landau and Lifshitz , e.g. see section 27 of Ref.[101], also
resort to a simple subtraction of the undesirable singularity at zero
temperatures. Hence, intuitive physical arguments may provide a reliable
guidance needed for further refinements of the results just presented.

\bigskip\ \ \ \ \ \ \ \ \ \ \ \ \ \ \ \ \ \ \ \ \ \ \ \ \ \ \ \ \ \ \ \ \ \
\ \ \ \ \ \ \ \ \ \ 

\textbf{Appendix A.2}

\ 

\textbf{Some results from the algebraic number theory}

\bigskip

Theory of arithmetic hyperbolic manifols has strong connections with the
algebraic number theory [102,103]. In this appendix we provide a condensed
summary of ideas on which such theory is based. Selection of topics is,
naturally, subjective and is meant only to encourage interested reader to
read much more comprehensive texts.

Development of number theory is associated with the desire to extend a
simple idea, known to nonprofessionals, that every nonnegative integer $n$
can be uniquely represented as 
\begin{equation}
n=p_{1}^{\alpha _{1}}\cdot \cdot \cdot p_{k}^{\alpha _{k}}  \tag{A.2.1}
\end{equation}
where $p_{i}$ is some prime number and $\alpha _{i}\geq 0$ is some integer.
This result becomes much less obvious if one is willing to extend it, say,
into domain of complex or algebraic numbers. For instance, 
\begin{equation}
6=2\cdot 3=(1+\sqrt{-5})(1-\sqrt{-5})=(4+\sqrt{10})(4-\sqrt{10}). 
\tag{A.2.2}
\end{equation}
Such non uniqueness is highly undesirable since many theorems of usual
arithmetic brake down. Solution of the nonuniqueness problem is one of the
major tasks of the modern number theory.

Let us define a \textbf{rational} number $\xi$ =p/q as a root of the
equation 
\begin{equation}
q\xi-p=0   \tag{A.2.3}
\end{equation}
with p and q being some integers. An \textbf{integer} is then a root of an
equation with first coefficient q=1. Analogously, an \textbf{algebraic}
number $\xi$ is any root of an algebraic equation 
\begin{equation}
a_{n}x^{n}+a_{n-1}x^{n-1}+\cdot\cdot\cdot+a_{0}=0   \tag{A.2.4}
\end{equation}
where the coefficients $a_{n},\cdot\cdot\cdot a_{0\text{ }}$ are rational
integers. Accordingly, an \textbf{algebraic integer }is any root of the
monic polynomial, e.g. see Eq.(3.22).

\bigskip

\textbf{Remark A 2.1}. Since only fibered knots and links are relevant for
dynamics of 2+1 gravity, e.g. read section 3, and since such knots/links are
associated with monic Alexander polynomials, we conclude, that the
arithmeticity is an intrinsic property of 2+1 gravity. Surely, much more is
associated with this observation as we shall demonstrate shortly.

\bigskip

In particular, let us consider the module (that is the set which is closed
under operations of addition and subtraction) of quadratic integers, i.e.
those, originating as roots of quadratic equation with $a_{2}=1.$ We had
discussed the roots of such equations already, e.g. see Eq.s(3.2), (3.9),
(3.18)and (3.35). Now we know that all these roots are \textbf{quadratic}
integers. As with usual (\textbf{rational})\textbf{\ integers}, it is
important to define the units, especially if we would like to go from the
module to a \textbf{ring} ( which is module where the multiplication
operation is defined which keeps all the members of the set within the set)
or to a \textbf{field} (where, in addition, the operation of division is
defined). It is clear that the typical representative of a quadratic integer 
$\xi $ should look like 
\begin{equation}
\xi =\frac{a+b\sqrt{d}}{c}  \tag{A.2.5}
\end{equation}
where $a,b,c$ and $d$ are some integers. Moreover, more detailed analysis
shows that $d$ can be chosen as square free. Let us notice, that in the case
if $d$=-1, we obtain more familiar field of complex numbers.Geometrically,
the integer lattice in such field can be built using the basis [1, i].
Therefore, in general case we can think of a basis [1, $\omega ]$ where 
\begin{equation}
\omega =\frac{1}{2}(d+\sqrt{d})  \tag{A.2.6}
\end{equation}
with $d$=D if D=2,3 or $d$=4D if D$\equiv $1 (mod 4) e.g. see Eq.(A.2.8) for
explanation of notations. The square free number $d$ is called \textbf{%
discriminant} of the field. This result confirms, in particular, that the
Markov triple in Eq.(5.22) is indeed made of integers since $\frac{1}{2}(3-%
\sqrt{-3})$ is just the conjugate of $\frac{1}{2}(3+\sqrt{-3}).$ For
different d's one will have different rings of quadratic integers.To make
rings, once the integers are defined, one should also define the units.This
can be done again by analogy with complex analysis. Using Eq.(A.2.5) we
define a norm $N(\xi )$ as 
\begin{equation}
N(\xi )=\xi \cdot \bar{\xi}  \tag{A.2.7}
\end{equation}
where $\bar{\xi}$ is conjugate of $\xi $, i.e. the same as $\xi $ in
Eq.(A.2.5) but with $b\rightarrow -b.$ In the number theory  the norm can be
both positive and negative.For quadratic imaginary field it is positive,
however. By definition, units are such numbers whose norm is unity. In the
case of usual complex numbers, one has therefore 4 units: $\pm 1$ and $\pm i$
, since in both cases the norm is one. In the case of $d$=-3 ring the units
are $\pm 1,\pm \rho ,\pm (1+\rho )$ and $\rho ^{2}$ where $\rho =\frac{1}{2}%
(-1+\sqrt{-3}),$in all other cases the number of units is 2, e.g. see
page156 of Ref.[102]$.$ The role of units can be easily understood by
analogy with more familiar case of ordinary integers . That is if $a$ and $b$
are integers then, $a=1\cdot b=b\cdot 1.$ Units play the same role for the
ring of quadratic numbers.These numbers are called \textbf{associates} if
they differ by unit factor. Next, one defines \textbf{primes} as those
quadratic numbers whose norm is a rational prime. In particular, we notice,
that the Markov triple, Eq.(5.22), is made of primes. Once the primes are
defined, one is naturally interested in finding an analogue of the
Eq.(A.2.1) for quadratic integers.This happens to be not a simple task: not
all d's will lead to unique decomposition into primes. For reasons which
will become clear upon reading of Appendix A.3, in the case of gravity only
negative d's are relevant. Among those which allow unique prime
decomposition are: -d=1, 2,3, 7, 11, 19,43, 67 and 163 [102,103] to be
compared with Eq.(5.23). Such comparison indicates, that some d's listed in
Eq.(5.23) suffer from non uniqueness. Such non uniqueness, had lead number
theoreticians to introduce the concepts of ideals and class numbers. Class
numbers reflect the extent to which integers of the field deviate from
uniqueness of decomposition. Naturally, if the decomposition is unique, the
class number is one. The ideals can be defined set theoretically and
geometrically as follows. Suppose,we have some set $I$ and a larger set $L$,
then $\forall $ $\alpha ,\beta \in I$ and $\forall \xi \in L$ we have 
\begin{equation*}
\alpha +\beta \in I\text{ \ \ (module property)}
\end{equation*}
and 
\begin{equation*}
\alpha \xi \in I\text{ (ideal property).}
\end{equation*}
Geometrically, this can be understood as follows. Consider, without loss of
generality, two dimensional lattice $\frak{L}$ built on vectors [1,$\omega ]$
then, for any rational integer n the point \ [n, n$\omega ]$ belongs to $%
\frak{L}$. Now one can think of different classes of rational integers and,
hence, of different sublattices $\frak{M}$ $\subset \frak{L}$ . This is
easily accomplished by introducing the residue classes modulo m. Recall,
that notation (equivalence relation) 
\begin{equation}
x\equiv y\text{ (}\func{mod}m\text{)}  \tag{A.2.8}
\end{equation}
means that x-y=mk where k is some integer.The above notation is just the
statement that when both x and y are divided by m they both have the same
residues. In particular, the statement x$\equiv 0(\func{mod}m)$ means that m
divides x. Let m have the same presentation as n in Eq.(A.2.1), then any
number x relatively prime to m may be determined ($\func{mod}m)$ by the
equations x$\equiv x_{i}(\func{mod}p_{i}^{a_{i}})$ , (x$_{i},p_{i})=1$, i=1,
2,..., k. The number of residue classes is just the Euler $\phi -$function $%
\phi (m).$ Now, if for a given sublattice $\frak{M}$ \ and vectors \textbf{x}
and \textbf{y} $\in \frak{L}$ we have equivalence relation 
\begin{equation}
\mathbf{x}\equiv \mathbf{y}\text{ (}\func{mod}\frak{M}),  \tag{A.2.9}
\end{equation}
this means only that \textbf{x}-\textbf{y}$\in \frak{M.}$ The number of
different sublattices within given lattice (i.e. the number of different
residue classes for ideals) is called index j. Evidently, j=[$\frak{L}$/$%
\frak{M}$]. \ Equivalently, it is also called a \textbf{norm} of an ideal $%
\frak{L}$ so that j=\textbf{N}[$\frak{L}$]. It happens, that in the case of
quadratic fields calculation of the norm is relatively easy [103]. If $a$ is
quadratic or rational integer, then \textbf{N}[$\frak{L}$]=$a^{2}$=$\left|
N(a)\right| $. For both ideals and integers the respective norms obey the
following composition law 
\begin{equation}
\text{\textbf{N}(}\frak{a}\text{)\textbf{N}(}\frak{b}\text{)=\textbf{N}(}%
\frak{ab}\text{)}  \tag{A.2.10}
\end{equation}
and 
\begin{equation}
N\text{(}\mathit{a}\text{)}N\text{(}\mathit{b}\text{)=}N\text{(}\mathit{ab}%
\text{).}  \tag{A.2.11}
\end{equation}
where the Gothic letters stand for the ideals and the italics for numbers.
The question arises: if there are prime numbers are there prime ideals? The
answer is ''Yes'' and since prime numbers play the central role in theory of
rational integers, one should expect that the prime ideals play the same
role in the number theory. This is indeed the case, but, unlike numbers,
every ideal is decomposable into prime ideals uniquely. This is the main
reason why these objects were introduced.

\bigskip

\textbf{Remark} \textbf{A 2.2.} In the case of ''quadratic fields'' there is
one to one correspondence between the ''quadratic ideals'' and the quadratic
forms[102 ]. We deliberately omit discussion of this fact to save the space
referring interested readers to literature \ [102-104].

\bigskip

Without going into details of such decomposition, we would like to take
advantage of the composition laws, Eqs. A.2.10-11.To this purpose we \ need
the following

\bigskip

\textbf{Theorem A 2.3.} \textit{The rational prime p factors in the
quadratic number}

\textit{\ field }$\mathbf{Q}(\sqrt{d})$\textit{\ (irrespective to the sign
of d), with accuracy up to}

\textit{\ multiplication by unit, as follows: }

\textit{a) p does not factor (p is \textbf{inert} in the field d) if and
only if (d/p)=-1,}

\textit{b) p \textbf{splits} in d into two different factors (e.g. z and z'
which belong to }

\textbf{Q}\textit{(}$\sqrt{d}$\textit{)}$)$\textit{\ if and only if \
(d/p)=-1,}

\textit{c) p is \textbf{ramified} in d (i.e. p=z}$^{2})$\textit{\ if and
only if (d/p)=0.}

\bigskip

In the above cases (d/p) is the Kronecker symbol which sometimes is denoted
as $\chi(p)$ for fixed d. $\chi(p)$ is actually a character (the Dirichlet's
character to be exact) of an Abelian group, e.g. of $\frak{L/M}$ . This
symbol should not be confused with $\delta_{ij}$ known in physics. At the
same time, the above theorem may serve as a definition of such a symbol. We
shall adopt this point of view in this work. Then, by analogy with the
theorem just stated, one can formulate the analogous theorem for the ideals
[103]

\bigskip

\textbf{Theorem A 2.4.} \textit{The quadratic prime ideals (}$\frak{p}$%
\textit{) are related to integers }$p$\textit{\ in the rational field in the
following possible ways:}

\textit{a) (}$\frak{p}$\textit{)=(}$\frak{p}$\textit{) , or (}$\frak{p}$%
\textit{) does not factor, i.e. inert, then \textbf{\ }}\textbf{N}\textit{(}$%
\frak{p}$\textit{)=}$p^{2};$

\textit{b) (}$\frak{p}$\textit{)=}$\frak{p}_{1}\frak{p}_{2}$\textit{, or (}$%
\frak{p}$\textit{) splits, then }\textbf{N}\textit{(}$\frak{p}_{1})$\textit{=%
}\textbf{N}\textit{(}$\frak{p}_{2})=p;$

\textit{c) (}$\frak{p}$\textit{)=}$\frak{p}_{1}^{2},$\textit{or (}$\frak{p}$%
\textit{) ramifies, then }\textbf{N}\textit{(}$\frak{p}_{1})=p$

\bigskip

Being armed with such results, we are ready to introduce the zeta function
for the field \ whose discriminant is $d$: 
\begin{equation}
\zeta _{d}(s)=\sum\limits_{\frak{a}}\left[ \text{N(}\frak{a}\text{)}\right]
^{-s}=\prod\limits_{\frak{p}}(1-\left[ \text{N(}\frak{p}\text{)}\right]
^{-s})^{-1}  \tag{A.2.11}
\end{equation}
where the last product is over all prime ideals. In view of the Theorem A
2.4., this result can be rewritten in more familiar terms as follows: 
\begin{equation}
\zeta _{d}(s)=\zeta (s)\sum\limits_{n=1}^{\infty }\frac{(d/n)}{n^{s}}\equiv
\zeta (s)L(s;d)  \tag{A.2.12}
\end{equation}
where $L$(s,d) is known in the literature as the Dirichlet L-series. It can
be shown [102-104], that such defined zeta function is having a pole
singularity only at s=1, i.e. the singularity only comes from the usual zeta
function. The residue of the pole at s=1 is much more interesting in the
present case as it is explained in the main text in section 5. Finally, we
need an ideal analogue of the Euler totient function $\phi (n).$ To this
purpose we take into account that an analogue of Eq.(A.2.1) is the same
equation in which all numbers are replaced by the ideals. Then, for an ideal 
$\frak{a}$ we have 
\begin{equation}
\Phi (\frak{a})=\text{\textbf{N}(}\frak{a}\text{)}\prod\limits_{\frak{p/a}%
}(1-\frac{1}{\text{\textbf{N}(}\frak{p}\text{)}}),  \tag{A.2.13}
\end{equation}
where $\frak{p}$ runs through the distinct prime divisors of $\frak{a}$ .

\bigskip

\ \textbf{Appendix A.3}

\bigskip

\textbf{Connections between the knot theory, theory of hyperbolic }

\textbf{spaces \ and the algebraic number theory: implications for }

\textbf{2+1 and \ 3+1 gravity}

\bigskip

For the sake of saving space and to avoid repetitions, we expect our readers
to be familiar with our earlier work, Ref.[25], while reading this Appendix.
Let us begin with the group PSL(2,C) of isometries of hyperbolic space H$^{3}
$ $.$ Only discrete subgroups of PSL(2,C) known as Kleinian groups are of
interest [98]. The group PSL(2,C) is just the projectivized version of \
SL(2,C). The following theorem for SL(2,C) is of particular interest, e.g.
see Ref.[98], chr.1,.

\bigskip

\textbf{Theorem A.3.1.} \textit{The group SL(2,C) is generated by two
elements} 
\begin{equation}
U_{1}=\left( 
\begin{array}{cc}
1 & a \\ 
0 & 1
\end{array}
\right) \text{ and }U_{2}=\left( 
\begin{array}{cc}
0 & -1 \\ 
1 & 0
\end{array}
\right) \text{ \ , where }a\in\mathbf{C.}   \tag{A.3.1}
\end{equation}

\bigskip

\textbf{Remark A.3.2.} This theorem is an analogue of the well known fact
[105] that the group PSL(2,Z) can be generated by the same two elements
(upon projectivization) with $a$=1. Hence, the parabolic elements play a
very special role in PSL(2,C) too.

\bigskip

The parabolic subgroups of PSL(2,C) are abelian subgroups associated with 
\textbf{Z} and \textbf{Z}$\oplus$\textbf{Z} cusps. If $\Gamma\in$PSL(2,C)
contains an abelian subgroup of rank one, i.e. \textbf{Z}-cusp, then the
associated fundamental domain for $\Gamma$ is non compact and of infinite
volume [29]. Hence, of no interest to us (additional reasons are given
below). If, however, $\Gamma\in$PSL(2,C) contains an abelian subgroup of
rank two, i.e. \textbf{Z}$\oplus$\textbf{Z} cusp, then the group $\Gamma$ is 
\textbf{not} cocompact but of \textbf{finite} covolume.\ The following
theorem was used in our earlier work, Ref.[25 ], and can be stated now as
follows :

\bigskip

\textbf{Theorem A.3.3.} \textit{Let} $\Gamma\in$PSL(2,C) \textit{be a
discrete group of finite covolume, then }$\Gamma$\textit{\ has only finitely
many classes of cusps.}

\bigskip

The proof can be found in Ref\textbf{.[}98\textbf{],} please, read Chr.2,
Proposition 3.8. .To insure that the subgroup $\Gamma$ is discrete the
following theorem is the most helpful [98]

\bigskip

\textbf{Theorem A.3.4}. (Jorgensen's inequality) \textit{If \TEXTsymbol{<}A,B%
\TEXTsymbol{>}is non-elementary discrete} \textit{subgroup of PSL(2,C),
then, the following inequality holds} 
\begin{equation}
\left| tr^{2}A-4\right| +\left| tr[A,B]-2\right| \geq 1.  \tag{A.3.2}
\end{equation}

\bigskip

\textbf{Remark A.3.5}. a) In the case of punctured torus extension of Markov
triples Eq.s(2.7),(2.8) to the field of quadratic integers was discussed in
Appendix A.2 and in the original works, Refs. [35,95]. Such an extension is
in accord with Jorgensen's inequality. b) This observation provides us with
an easy way of definig the arithmetic Klenian groups and, hence, the
arithmetic 3-manifolds. To this purpose, we need still one more theorem [95].

\bigskip

\textbf{Theorem A.3.6.} \textit{\ Let }$A_{1},A_{2},...,A_{n}\in $\textit{\
SL(2,C) be such that }$A_{1}$\textit{\ is not parabolic and} \TEXTsymbol{<}$%
A_{1},A_{2}>$ \textit{is not elementary. Then, the }$A_{1},A_{2},...,A_{n}$%
\textit{\ are determined up to conjugacy in} SL(2,C) \textit{by} $Tr$ $A_{i}$%
, $TrA_{i}A_{j}$ \textit{and} $TrA_{i}A_{j}A_{k}$ \textit{with} i,j,k $%
\in\{1,2,...,n\}.$

\bigskip

\textbf{Remark A.3.7}. The \textbf{arithmetic} Kleinian groups can be
regarded as images in PSL(2,C) of subgroups of SL(2,C) so that the trace of
an element in the Kleinian group is defined up to a sign. The traces $Tr$ $%
A_{i}$ belong to the field of quadratic integers with negative discriminant
d. The negativity of d is explained shortly below.

\bigskip

Following [106], page 192, Theorem E.4.4.,and also Riley [29] consider
complement of a link $\mathcal{L}$ in S$^{3}$ (which is just a union of
knots $\mathcal{L}_{1},...,\mathcal{L}_{k})$.To this purpose, let T$_{i}$ be
an open tubular neighborhood of the knot $\mathcal{L}_{i}$ (that is T$_{i}$
is a solid torus with boundary removed). Then, the following theorem holds:

\bigskip

\textbf{Theorem A.3.8}. \textit{Given a compact 3-manifold N there exists a
link }$\mathcal{L}$\textit{\ in S}$^{3}$\textit{\ which is} \textit{union of
knots} $\mathcal{L}_{1},...,\mathcal{L}_{k}$, \textit{such that N is
obtained from the manifold} 
\begin{equation*}
M=S^{3}\setminus (\bigcup\limits_{i=1,...,k}\text{T}_{i}),
\end{equation*}
\textit{where }T$_{i}^{\prime }s$\textit{\ are pairvise disjoint open
tubular neighborhoods of the knots }$\mathcal{L}_{i},$\textit{by the
operation} \textit{of Dehn filling.}

\bigskip

\textbf{Corollary A.3.9}. (Riley and Thurston [19,29]) \textit{If }$M$%
\textit{\ is to be hyperbolic, then every non-cyclic abelian subgroup (of
rank two) of }$\pi_{1}M$\textit{\ must be peripheral and correspond to the
component }$\mathcal{L}_{i\text{ }}$\textit{of link }$\mathcal{L}.$

\bigskip

\textbf{Remark A.3.10.} Since the fundamental group $\pi_{1}M$ can be
embedded into PSL(2,C) and since the non-cyclic peripheral (e.g.section 2
for definition) abelian subgroup of rank two corresponds to \textbf{Z}$%
\oplus $\textbf{Z} cusp there is a one to one correspondence between the
cusps and the link components. E.g. in the case of figure eight knot we have
hyperbolic 3-manifold with just one cusp [19,29]. Moreover, and this is the
most important, the following theorem of Margulis [30] requires such cusped
3-manifolds to be arithmetic.

\bigskip

\textbf{Theorem A.3.11}.(G.Margulis) \textit{Let S be Riemannian symmetric
noncompact space of rank greater than one and }$\Lambda $\textit{\ is
irreducible discrete group of motions of S such that S/}$\Lambda $\textit{\
has finite volume but noncompact. Then, }$\Lambda $\textit{\ is the
arithmetic subgroup in the group of motions of space S.}

\bigskip

\textbf{Definition A.3.12 }( Helgason [31], Chr.5 ) Let S be the Riemannian
globally symmetric space. The \textbf{rank} of S is the maximal dimension of
a flat, \textbf{totally geodesic} subspace M. The subspace M is geodesic at $%
p\in M$ if each S-geodesic which is tangent to M at $p$ is a curve in M. The
submanifold \ M is totally geodesic \ if it is geodesic at each of its
points.

\bigskip

\textbf{Remark} \textbf{A.3.13}. To use the above definition, the following
alternative interpretation of rank is helpful. Following Besse [32], let $%
\frak{g}$=$\frak{h}$+$\frak{m}$ be decomposition (into irreducible parts) of
the Lie algebra of the group of motions of space S where $\frak{h}$ is non
and $\frak{m}$ is abelian parts.Then, the rank is the total dimension of the
abelian subalgebra $\frak{m}$ of $\frak{g}$.

\bigskip

\textbf{Remark A.3.14}. Since H$^{3}$ is symmetric space, Margulis theorem
implies arithmeticity of hyperbolic 3-manifolds with \textbf{Z}$\oplus $%
\textbf{Z} cusps. In the case of one cusp the complement of figure eight
knot hyperbolic 3-manifold is indeed arithmetic as it was shown
independently by Reid [97]. The arithmeticity can actually be anticipated
based on the results of our section 3. Indeed, since surface automorphisms
are associated with the roots of the monic Alexander polynomial whose
coefficients are rational integers, these roots lie necessarily in the field
of quadratic integers. The Seifert-fibered phase is associated with units of
the field while the pseudo-Anosov with nonunit integers.Since these stretch
factors are related to the traces of the corresponding matrices, e.g. see
Eq.(2.15), connection with Theorem A.3.6. is clear.

\bigskip

\textbf{Remark A.3.15}. Since Margulis theorem is not limited to groups of
motions of H$^{3}$ its application can be extended to 3+1 gravity. According
to Besse,Chr 10 of Ref.[32], every irreducible symmetric space is Einstein
space.The classification of irreducible symmetric spaces was made by
Helgason [31] and is also cited in Besse's book. From this classification it
follows that most of such spaces have rank higher than one and, hence,
should be arithmetic. It would be interesting to know if there is an
analogue of Theorem A.3.3 for Einstein spaces. This is of interest because,
from the discussion we had in section 5, it follows that the cusps act very
much like black holes, e.g. see Fig. 11, (except in 2+1 case the
''particle'' cannot be trapped by the ''black hole'') and, therefore, the
role of black holes in the Universe acquires new meaning: they make our
Universe arithmetic. If this is the case, then the Universe becomes
something like a crystalline solid which potentially can undergo phase
transitions between lattices of different symmetry. The symmetry of the
lattice is determined by the discriminant of the quadratic number field.

\bigskip

\textbf{Remark A.3.16}. The connection between the cusps and the black holes
is not purely visual, e.g. Fig.11, or formal. It is real, at least for 2+1
Euclidean black hole ! Indeed, following Carlip, e.g. see p. 50 of Ref[22],
by the appropriate choice of coordinates and units the metric of the
Euclidean black hole becomes that of H$^{3},$ i.e. 
\begin{equation*}
\left( ds\right) ^{2}=\frac{\left( dx\right) ^{2}+\left( dy\right)
^{2}+\left( dz\right) ^{2}}{z^{2}}, 
\end{equation*}
provided that \ in addition the following identification is made, 
\begin{equation*}
(x,y,z)\sim e^{2\pi r_{+}}(x\cosh2\pi r_{+}-y\sinh2\pi r_{+},\text{ }%
y\cosh2\pi r_{+}-x\sin nh2\pi r_{+}\text{, }z\text{),}
\end{equation*}
which is easily recognizable as \textbf{Z}$\oplus$\textbf{Z} cusp [25].

\bigskip

\textbf{Remark A.3.17}. The reasons for existence of black holes in nature
were discussed in the past [107 ]. Ironically, the same author had came to
the opposite conclusions in [108], that is the black hole singularity is
fictitious.The same conclusion is reached in Ref.[109] which claims that two
dimensional quantum black holes are nonsingular.

\bigskip

Now we have to explain why this discriminant should be negative. For one
thing, if it would be positive the norm of the quadratic field would be
negative. This is unphysical in view of Remark A.3.14. There are other
reasons in addition which we would like now to explain. Let us begin with
famous example by Riley [110] who considered the mapping of the fundamental
group $\pi_{8}$ of the complement of figure eight knot into PSL(2,C).The
fundamental group $\pi_{8}$ is free non abelian group composed of two
generators which has the following presentation 
\begin{equation}
\pi_{8}=<x_{1},x_{2};wx_{1}=x_{2}w>   \tag{A.3.3}
\end{equation}
where $w=x_{1}^{-1}x_{2}x_{1}x_{2}^{-1}.$ The homomorphism of $\pi_{8}$ into
PSL(2,C) can be achieved with help of the Theorem A.3.1.as follows: 
\begin{equation}
x_{1}\rightarrow A=\left( 
\begin{array}{cc}
1 & 1 \\ 
0 & 1
\end{array}
\right) \text{ and }x_{2}\rightarrow B=\left( 
\begin{array}{cc}
1 & 0 \\ 
-\omega & 1
\end{array}
\right)   \tag{A.3.4}
\end{equation}
where $\omega$ is some number, different from 1 and to be determined
below.The matrix $B$ is just a conjugate of $U_{2}$ in Eq.(A.3.1), i.e.$%
B=U_{1}U_{2}U_{1}^{-1}$ with $a\leftrightarrows\omega.$ Since these matrices
should obey the relation $wx_{1}=x_{2}w$, replacements $x_{1}\rightarrow A$
and $x_{2}\rightarrow B$ produce the following equation for $\omega$%
\begin{equation}
\omega^{2}+\omega+1=0   \tag{A.3.5}
\end{equation}
with solution $\omega=\frac{1}{2}(-1+\sqrt{-3}).$ One can easily recognize
the unit $\rho$ of the quadratic field with discriminant 3 (from Appendix
A.2). Hence, figure 8 knot is indeed arithmetic. From this example it should
be clear, that the arithmeticity is associated with our choice of $\omega.$

Consider now the subgroup of PSL(2,C) which fixes j (i.e. in the upper space
model H$^{3}$ with coordinates z+jt (z=x+iy, t\TEXTsymbol{>}0) we are
looking for the M\"{o}bius transformation M(j)=j). It is easy to prove [98]
that such transformation is associated with SL(2,C) matrix of the type 
\begin{equation}
M=\left( 
\begin{array}{cc}
x & y \\ 
-\bar{y} & \bar{x}
\end{array}
\right)   \tag{A.3.6}
\end{equation}
such that $\det M=\left| x\right| ^{2}+\left| y\right| ^{2}=1$ where $\left|
x\right| ^{2}=x\bar{x}$ and \={x} is a complex conjugate of x, etc. Such
matrix $M$ is in one to one correspondence with the field $\mathcal{H}$ of
quaternions q known in physics literature as Hamiltonian quaternions [111] .
Such quaternions usually are represented as q=a +bi +cj +dk where a,b,c,d
are real numbers and k=ij. In terms of quaternions Eq. (A.3.6) can be
written as 
\begin{equation}
M(q,\mathcal{H})=\left( 
\begin{array}{cc}
a+bi & c+di \\ 
-c+di & a-bi
\end{array}
\right) .  \tag{A.3.7}
\end{equation}
Surely, if we write i=$\sqrt{-1}$ , then the above matrix represents the
most general matrix associated with the unit of the imaginary quadratic
field with discriminant d=-1. Naturally, one can therefore think of the most
general matrix associated with units of imaginary quadratic fields with
discriminant d \textbf{other} than -1.To this purpose one needs to extend
the notion of quaternions known in physics.

Let K be the number field of characteristic different from 2 (the
characteristic p of the field K is associated with subdivision of the field
into residue classes Z$_{p}$ , so p$\neq2$ and p is prime$)$ and let $a,b\in$%
K be two non-zero elements. A quaternion algebra $\mathcal{H}$($a$,$b$;K)
over K is generated by elements $i$ and $j$ satisfying $i^{2}=a,$ $j^{2}=b$
and $ij=-ij$. The elements 1, $i,j,ij$ form a K-basis of $\mathcal{H}$($a$,$b
$;K) as a vector space. The pair (a,b) is called the Hilbert symbol for $%
\mathcal{H}$ . In particular, for Hamiltonian quaternions the field is 
\textbf{R }and the Hilbert symbol is (-1,-1). If $L$ is the field extension
of K with $\sqrt{a}$ and $\sqrt{b}$ $\in L$ then, repeating all the steps
leading to Eq.(A.3.7), we obtain, 
\begin{equation}
M(2,L)=\left( 
\begin{array}{cc}
x_{0}+x_{1}\sqrt{a} & x_{2}\sqrt{b}+x_{3}\sqrt{ab} \\ 
x_{2}\sqrt{b}-x_{3}\sqrt{ab} & x_{0}-x_{1}\sqrt{a}
\end{array}
\right) .   \tag{A.3.8}
\end{equation}
Newman and Reid [112] proved the following major

\bigskip

\textbf{Theorem A.3.18.} \textit{Let }$\Gamma$\textit{\ be a non-compact
Kleinian group of finite covolume.Then:}

\textit{a) }$\Gamma$\textit{\ is arithmetic if and only if the trace field
(e.g. see Theorem A.3.6) }K\textit{=\textbf{Q}(}$\sqrt{-d}$\textit{) for
some square free d}$\in N$\textit{\ (e.g.see Eq.5.23) and }$\mathit{tr}%
\Gamma $\textit{\ consists of algebraic integers.}

\textit{b) }$\Gamma$\textit{\ is derived from the quaternion algebra }$%
\mathcal{H}$\textit{(}$a$\textit{,}$b$\textit{;K) if and only if }$\mathit{tr%
}\Gamma\subset\mathcal{O}_{d\text{ }}$\textit{\ for some d.}

\bigskip

\pagebreak

\bigskip

\ \ \ \ \ \ \ \ \ \ \ \ \ \ \ \ \ \ \ \ \ \ \ \ \ \ \ \ \ \ \ \ \ \ \ \ \ \
\ \ \ \ \ 

\ \ \ \ \ \ \ \ \ \ \ \ \ \ \ \ \ \ \ \ \ \ \ \ \ \ \ \ \ \ \ \ \ \ \ \ \ \
\ \ \ \ \ \ \ \textbf{\ References}

\bigskip

[1] E.Titchmarsh, The Theory of the Riemann Zeta-Function,

\ \ \ Clarendon Press, Oxford, 1986.

[2] H. Edwards, Riemann's Zeta Function, Academic Press, New York, 1974.

[3] S.Patterson, An Introduction to the Theory of the Riemann Zeta

\ \ \ \ \ Function, Cambridge U.Press,1988.

[4]\ M.Berry and J.Keating, The Riemann zeros and eigenvalue asymptotics,

\ \ \ \ \ SIAM Reviews 41 (1999) 236-266.

[5] A.Connes, Trace formula in noncommutative geometry and zeros of the

\ \ \ \ Riemann zeta function, Sel.Math.New Series 5 (1999) 23-106.

[7] C.Yang and T.Lee, Statistical theory of equations of state and phase

\ \ \ transitions,\ Phys.Rev.87 (1952) 404-419.

[8] A.Knauf, On a ferromagnetic spin chain,Comm.Math.Phys.

\ \ \ \ 153 (1993) 77-115.

[9] A.Kontucchi and A.Knauf, The phase transition of the number-theoretic

\ \ \ \ \ spin chain, Forum Mathematicum, 9 (1997) 547-567.

[10] A.Knauf, Phases of the number-theoretic spin chain, J.of Statistical

\ \ \ \ \ \ Physics 73 (1993) 423-431.

[11] B.Pavlov and L.Faddeev, Scattering theory and automorphic functions,

\ \ \ \ \ \ \ J.Sov.Math. 3 (1975) 522-548.

[12] P.Lax and R.Phillips, Scattering Theory for Automorphic Functions,

\ \ \ \ \ \ \ Princeton U.Press, Princeton, 1976.

[13] \ M.Gutzwiller, Stochasic behavior in quantum scattering ,

\ \ \ \ \ \ \ PhysicaD 7 (1983) 341-355.

[14] R.Artuso, P.Cvitanovic and B.Kenny, Phase transitions on strange

\ \ \ \ \ \ irrational sets,\ Phys.Rev. A 39 (1989) 268-281.

[15] P.Cvitanovic, Circle maps: Irrationality Winding, in From Number

\ \ \ \ \ Theory to Physics , M.Waldschmidt, P.Moussa, J.-M.Luck and

\ \ \ \ \ C.Itzykson eds., Springer-Verlag, Berlin 1992.

[16] A.Kholodenko, Use of meanders and train tracks for description

\ \ \ \ \ \ of defects and textures in liquid crystals and 2+1 gravity, J.of
Geom.

\ \ \ \ \ \ and Phys.33(2000) 23-58.

[17] A.Kholodenko, Use of quadratic differentials for description of

\ \ \ \ \ \ defects and textures in liquid crystals and 2+1 gravity, J. of
Geometry

\ \ \ \ \ and Physics 33 (2000) 59-102.

[18] \ G.Brude and H.Ziechang, Knots, Walter de Gruyter, Berlin, 1985.

[19] \ W.Thurston, Geometry and Topology of 3-Manifolds, Princeton U.Lecture

\ \ \ \ \ \ \ \ Notes, 1979 (http://www.msri.org/gt3m/)

[20] P.Chaikin and T.Lubensky,Principles of Condensed Matter Physics,

\ \ \ \ \ \ \ Cambridge University Press, Cambridge, 1995.

[21] C.Godbillon, Feuilletages, Birkhauser, Boston, 1991.

[22] S.Carlip, Quantum gravity in 2+1 dimensions, Cambridge University

\ \ \ \ \ \ \ Press, Cambridge, 1998.

[23] J.Kosterlitz and D.Thouless, Metastability and phase transitions in

\ \ \ \ \ \ \ \ two dimensional systems, J.Phys.C6 (1973) 1181-1203.

[24] J.Peyriere, Trace maps in \ Beyond Quasicrystals, pp.465-480,

\ \ \ \ \ \ Springer-Verlag, Berlin, 1995.

[25] A.Kholodenko, Boundary conformal field theories, limit sets of Kleinian

\ \ \ \ \ \ groups and holography, J.of Geom.and Phys.(2000) \ 1-46 \ (in
press).

[26] P.Sarnak, The arithmetic and geometry of some hyperbolic three

\ \ \ \ \ \ \ manifolds, Acta Math.151 (1983) 253-295.

[27] \ I.Efrat and P.Sarnak, The determinant of the Eisenstein matrix and

\ \ \ \ \ \ \ Hilbert class fields, AMS Transactions 290 91985) 815-824.

[28] \ L.Bianchi, Sui gruppi de sostituzioni lineari con coefficienti

\ \ \ \ \ \ \ \ appartenenti a corpi quadratici immaginari, Math. Ann.

\ \ \ \ \ \ \ 40 (1892) 332-412.

[29] R.Riley, An elliptical path from parabolic representations to

\ \ \ \ \ \ \ hyperbolic structures, LNM 722 (1979) 99-133.

[30] G.Margulis, Arithmetic properties of discrete groups, Russian Math.

\ \ \ \ \ \ \ Surveys 29 (1974) 49-98.

[31] S.Helgason, Differential Geometry, Lie Groups, and Symmetric Spaces,

\ \ \ \ \ \ \ \ Academic Press, New York, 1978.

[32] \ A.Besse, Einstein Manifolds, Springer Verlag, Berlin, 1987.

[33] \ J.Libre and R.Mackey, Pseudo-Anosov homeomorphisms on a sphere

\ \ \ \ \ \ \ with\ four punctures have all periods, Math.Proc.Cambr.Phil.
Soc.112

\ \ \ \ \ \ \ \ \ (1992) 539-549.

[34] R.Penner, An introduction to train tracks in LMS Lecrure Notes Series

\ \ \ \ \ \ \ 112 (1986) 77-90.

[35] B.Bowdich, Markov triples and quasifuchsian groups, Proc.London Math.

\ \ \ \ \ \ \ Soc.77 (1998) 697-736.

[36] T.Cusick and M.Flahive, The Markov and Lagrange Spectra,

\ \ \ \ \ \ AMS Publishing, Providence, 1989.

[37] B.Bowdich, A variation of McShane's identity for once punctures

\ \ \ \ \ \ \ torus bundles, Topology 36(1997) 325-334.

[38] T.Jorgensen, On pairs of punctured tori, unpublished manuscript,

\ \ \ \ \ \ \ Columbia University.

[39] H.Cohn, Approach to Markoff's minimal forms through modular

\ \ \ \ \ \ functions, Ann.Math.61 (1955) 1-12.

[40] J.Roberts, Escaping orbits and trace maps, Physica\thinspace A

\ \ \ \ \ \ \ 228 (1996) 295-325.

[41] Y.Imayoshi and M.Taniguchi, An Introduction to Teichm\"{u}ller Spaces,

\ \ \ \ \ \ \ Springer-Verlag, Berlin, 1992.

[42] R.Penner, Bounds on the least dilatations, AMS Proceedings 113 (1991)

\ \ \ \ \ \ \ 443-450.

[43] A.Haas, Diophantine approximation on hyperbolic Riemann surfaces,

\ \ \ \ \ \ Acta Math. 156 (1986) 33-82.

[44] R.Alperin, W.Dicks and J.Porti, The boundary of the Giseking

\ \ \ \ \ \ tree in hyperbolic three space, Topology and its Applications

\ \ \ \ \ 93 (1999) 219-259.

[45] G.Hardy and E.Wright, An introduction to the theory of numbers,

\ \ \ \ \ \ \ Clarendon Press, Oxford, 1962.

[46] C.Series,The modular surface and continued fractions, J.London Math.Soc.

\ \ \ \ \ \ \ 31(1985) 69-80.

[47] Y.Minsky, The classification of punctured torus groups,
Ann.Math.149(1999)

\ \ \ \ \ \ \ 559-626.

[48] H.Rademacher, Higher Mathematics From an Elentary Point of View,

\ \ \ \ \ \ \ Birkh\"{a}user, Boston, 1983.

[49] H.Cohn, Mathematical microcosm of geodesics, free groups and Markoff

\ \ \ \ \ \ \ forms, LNPAM 149 (1999) 559-626.

[50] A.Beardon, J.Lehner and M.Sheingorn, AMS Transactions 295 (1986)

\ \ \ \ \ \ 635-647.

[52] R.Courant and H.Robbins, What is Mathematics?, Oxford

\ \ \ \ \ \ \ University Press, Oxford, 1996.

[53] R.Penner and J.Harer, Combinatorics of Train Tracks,

\ \ \ \ \ \ Princeton University Press, 1992.

[53] W.Lok, Deformation of Locally Homogenous Spaces and Kleinian Groups,

\ \ \ \ \ \ \ \ PhD Thesis, Columbia University, 1984.

[54] N.Gilbert and T.Porter, Knots and Surfaces, Oxford University

\ \ \ \ \ \ \ Press, Oxford, 1994.

[55] E.Witten, 2+1 dimensional gravity as exactly soluble problem,

\ \ \ \ \ \ \ Nucl.Phys.B311 (1988) 46-78.

[56] C.McMullen, Renormalization and 3-Manifolds which Fiber

\ \ \ \ \ \ Over the Circle, Princeton University Press, Princeton, 1996.

[57] A.Kholodenko and Th.Vilgis, Some geometrical and topological

\ \ \ \ \ \ \ problems in polymer physics, Phys.Reports 298 (1998) 251-370.

[58] D.Rolfsen, Knots and Links, Publish or Perish, Houston, 1990.

[59] D.Collins, R.Grigorchuk,P.Kurchanov and H.Zieschang, Combinatorial

\ \ \ \ \ \ \ Group Theory and Applications to Geommetry, Springer-Verlag,

\ \ \ \ \ \ \ Berlin, 1998.

[60] W.Thurston,Three-Dimensional Geometry and Topology,

\ \ \ \ \ \ \ Princeton University Press, Princeton, 1997.

[61] J-P Otal, Le theoreme d'hyperbolization pour les varietes fibres

\ \ \ \ \ \ de dimension 3, Asterisque 235 (1996) 1-159.

[62] H.Morton, Fibered knots with given Alexander polynomial,

\ \ \ \ \ \ \ Enseignment Math.31(1983) 205-222.

[63] H.Matschul, On the relation between 2+1 Einstein gravity

\ \ \ \ \ \ \ and Chern-Simons theory, Class. Quantum Gravity

\ \ \ \ \ \ \ 16 (1999) 2599-2609.

[64] E.Rykken, Markov partitions and the expanding factor for pseudo-

\ \ \ \ \ \ \ Anosov homeomorphisms, PhD Thesis, Nortwestern University,

\ \ \ \ \ \ \ 1993.

[65] W.Jaco and P.Shalen, Seifert fibered spaces in 3-manifolds,

\ \ \ \ \ \ \ AMS Memoirs 21 (1979) 1-292.

[66] P.Orlik, Seifert Manifolds, LNM 291 (1972) 1-155.

[67] R.Kulkarni, K. Lee and F.Raymond, Deformation spaces for

\ \ \ \ \ \ Sefert manifolds, LNM 1167 (1985) 180-216.

[68] G.Everest and Th.Ward, Heights of Polynomials and Entropy in

\ \ \ \ \ \ \ Algebraic Dynamics, Springer-Verlag, Berlin, 1999.

[69] P.Walters, An Introduction to Ergodic Theory,

\ \ \ \ \ \ \ Springer-Verlag, Berlin,1982.

[70] U.Oerttel, Incompressible branched surfaces, Inventiones

\ \ \ \ \ \ \ Math.76(1984) 385-410.

[71] W.Floyd and U.Oertel, Topology 23 91984) 117-125.

[72] A.Kawauchi, A Survey of Knot Theory, Birhauser,

\ \ \ \ \ \ \ Boston, 1996.

[73] M.Culler, W.Jaco and H.Rubinstein, Incompressible surfaces

\ \ \ \ \ \ \ in once punctured torus bundles, Proc.London.Math.Soc.45

\ \ \ \ \ \ \ (1982) 385-419.

[74] W.Floyd and A.Hatcher, Incompressible surfaces in punctured

\ \ \ \ \ \ \ torus bundles, Topology and its Appl. 13 (1982) 263-282.

[75] W.Jaco, Lectures on Three-Manifold Topology, AMS,

\ \ \ \ \ \ \ \ Providence, 1980.

[76] W.Neumann, Notes on geometry and 3-manifolds in Low

\ \ \ \ \ \ \ \ Dimensional Topology, J.Bolyai Math.Soc., Budapest, 1999.

[77] A.Hatcher, On the boundary curves of incompressible surfaces,

\ \ \ \ \ \ \ Pacific Journal of Mathematics 99(1982) 373-377.

[78] S.Lang, Introduction to Diophantine Approximations,

\ \ \ \ \ \ \ Springer-Verlag, Berlin, 1995.

[79] A.Kholodenko, Fermi-Bose transmutation: from semiflexible

\ \ \ \ \ \ polymers to superstrings, Ann.Phys.202 (1990) 186-225.

[80] J.Przytycki, Incompressible surfaces in 3-manifolds,

\ \ \ \ \ \ \ PhD Thesis, Columbia University, 1981.

[81] J.Przytycki, Incompressibility of surfaces after Dehn surgery,

\ \ \ \ \ \ Michigan Math.Journ.30 (1983) 289-308.

[82] P.Contucci,P.Kleban and A.Knauf, A fully magnetizing phase

\ \ \ \ \ \ transition, J.Stat.Phys. 97 (1999) 523-539.

[83] P.Kleban and A.Qzluk, A Farey fraction spin chain,

\ \ \ \ \ \ \ Comm.Math.Phys. 203 (1999) 635-647.

[84] R.Penner, The universal Ptolemy group and its completions,

\ \ \ \ \ \ \ London Math.Soc.Lect.Notes Series 243 (1997) 293-312.

[85] B.Bowditch, A proof of McShanes identity via Markoff

\ \ \ \ \ \ \ triples, Bull.London Math.Soc. 28 (1996) 73-78.

[86] G.McShane, A remarkable identity for lerngths of curves,

\ \ \ \ \ \ \ PhD Thesis, University of Warwick, 1991.

[87] S.Katok, Fuchsian Groups,The University of Chicago

\ \ \ \ \ \ \ Press, Chicago, 1992.

[88] D.Sullivan, Disjoint spheres, approximation by imaginary quadratic

\ \ \ \ \ \ numbers, and the logarithm law for geodesics, Acta Math.

\ \ \ \ \ \ 149 (1982) 215-237.

[89] S.Cosentino, Equidistribution of parabolic fixed points in the

\ \ \ \ \ \ \ limit set of Kleinian groups, Ergodic Th. and Dyn.Systems 19

\ \ \ \ \ \ (1999)1437-1484.

[90] D.Sullivan, Entropy, Hausdorff measures old and new, and limit

\ \ \ \ \ \ sets of geometrically finite Kleinian groups, Acta Math.

\ \ \ \ \ \ 153(1984) 259-277.

[91] A.Beardon, The Geometry of Discrete Groups, Springer-Verlag,

\ \ \ \ \ \ \ New York, 1983.

[92] Y.Motohashi, Spectral Theory of the Riemann Zeta Function,

\ \ \ \ \ \ \ Cambridge University Press, Cambridge, 1997.

[93] A.Venkov, Spectral theory of automorphic functions,

\ \ \ \ \ \ Selberg's Zeta function and some problems of analytical

\ \ \ \ \ \ \ number theory and mathematical physics, Russian Math.Surveys

\ \ \ \ \ \ \ 34 (1979) 69-135.

[94] W.Neumann and D.Zagier, Volumes of hyperbolic three manifolds,

\ \ \ \ \ \ \ \ Topology 24 (1985) 307-332.

[95] B.Bowditch, C.McLachlan and A.Reid, Arithmetic hyprbolic

\ \ \ \ \ \ \ surface bundles, Math.Ann.302 (1995) 31-60.

[96] F.Grunewald and U.Hirsch, Link complements arising from

\ \ \ \ \ \ \ arithmetic group actions, Int.Journ.of Math. 6 (1995) 337-370.

[97] A.Reid, Arithmeticity of knot complements, J.London Math.Soc.

\ \ \ \ \ \ 43 (1991) 171-184.

[98] J.Elstrodt, F.Grunewald and J.Mennicke, Groups Acting on

\ \ \ \ \ \ Hyperbolic Space, Springer-Verlag, Berlin, 1998.

[99] K.Huang, Statistical Physics, John Wiley \& Sons Inc.,

\ \ \ \ \ \ New York, 1963.

[100] R.Feynman, Statistical Mechanics, Addison-Wesley Publising

\ \ \ \ \ \ \ \ \ Co., Reading, MA, 1990.

[101] L.Landau and L.Livshits, Statistical Mechanics, Nauka,

\ \ \ \ \ \ \ \ \ Moscow, 1976.

[102] E.Hecke, Lectures on the Theory of Algebraic Numbers,

\ \ \ \ \ \ \ \ \ Springer-Verlag, Berlin, 1981.

[103] H.Cohn, Advanced Number Theory, Dover,

\ \ \ \ \ \ \ \ \ New York,1980.

[104] H.Weil, Algebraic Theory of Numbers, Princeton University

\ \ \ \ \ \ \ \ \ Press, Princeton, 1940.

[105] H.McKean and V.Moll, Elliptic Curves, Cambridge

\ \ \ \ \ \ \ \ \ University Press, Cambridge, 1999.

[106] R.Benedetti and C.Petronio, Lectures on Hyperbolic

\ \ \ \ \ \ \ \ \ Geometry, Springer-Verlag, Berlin, 1992.

[107] G.Horowitz and M.Perry, Black holes and the stability of

\ \ \ \ \ \ \ \ \ gravitation, General Relativity and Gravitation 15 (1983)
1-5.

[108] V.Kostelesky and M.Perry, No more spacetime singularities,

\ \ \ \ \ \ \ \ \ General Realativity and Gravitation 26 (1994) 7-12.

[109] E.Teo, What have we learned from two dimensional models

\ \ \ \ \ \ \ \ \ of quantum black holes?, General Relativity and
Gravitation,

\ \ \ \ \ \ \ \ \ \ 26 (1994) 13-19.

[110] A.Riley, A quadratic parabolic group, Math.Soc. Cambridge

\ \ \ \ \ \ \ \ Phil.Soc.77 (1975) 281-288.

[111] C.Misner, K.Thorne and J.Wheeler, Gravitation,

\ \ \ \ \ \ \ \ \ W.H.Freeman and Co., San Francisco, 1973.

[112] W.Neumann and A.Reid, Arithmetic of hyperbolic manifolds in

\ \ \ \ \ \ \ \ \ \ Topology 90, pp 273-310, De Gyiter-Verlag, Berlin, 1992.

\ \ 

\ \ \ \ \ \ 

\ \ \ \ \ 

\ \ \ \ \ \ 

\ \ \ \ \ \ 

\ \ \ \ \ 

\ \ \ 

\ \ \ 

\bigskip

\end{document}